\documentclass[twocolumn]{aastex701}
\pdfoutput=1 
\usepackage{amsmath,amstext}
\usepackage[T1]{fontenc}
\usepackage{apjfonts} 
\usepackage[figure,figure*]{hypcap}
\usepackage{url}
\usepackage{CJKutf8}
\usepackage{array} 
\newcommand{\cellbox}[2]{\parbox[t]{#1}{\raggedright #2}}

\shorttitle{Critical Metallicity}
\shortauthors{Ou et al.}

\begin{document}
\begin{CJK*}{UTF8}{bsmi}
\title{Critical Metallicity of Cool Supergiant Formation. II. Physical Origin}

\author[0000-0003-1295-8235]{Po-Sheng Ou （歐柏昇）}
\affiliation{Institute of Astronomy and Astrophysics, Academia Sinica, No.1, Sec. 4, Roosevelt Rd., Taipei 106319, Taiwan, R.O.C.} 
\affiliation{Department of Physics, National Taiwan University, No.1, Sec. 4, Roosevelt Rd.,  Taipei 106319, Taiwan, R.O.C.} 
\email{psou@asiaa.sinica.edu.tw}

\correspondingauthor{Po-Sheng Ou}
\email{psou@asiaa.sinica.edu.tw}

\author[0000-0002-4848-5508]{Ke-Jung Chen（陳科榮）}
\affiliation{Institute of Astronomy and Astrophysics, Academia Sinica, No.1, Sec. 4, Roosevelt Rd., Taipei 106319, Taiwan, R.O.C.} 
\affiliation{Heidelberg Institute for Theoretical Studies, Schloss-Wolfsbrunnenweg 35, 
Heidelberg 69118,
Germany}
\email{kjchen@asiaa.sinica.edu.tw}

\begin{abstract}
This study investigates the physical origin of the critical metallicity required for the formation of cool supergiants, as revealed by stellar evolution models. Using grids of stellar models, we show that the terminal-age main-sequence (TAMS) radius, $R_{\rm TAMS}$, defines a threshold that determines whether a star of a given mass can evolve into the red supergiant (RSG) phase. Metallicity influences the supergiant outcome because it modifies $R_{\rm TAMS}$ through its effects on opacity and nuclear energy generation, as demonstrated by our stellar models and dimensional analysis based on homology relations. The value of $R_{\rm TAMS}$ sets the initial radius for post-main-sequence expansion and therefore controls the envelope radius reached at subsequent core-evolution stages. Higher-metallicity stars develop larger $R_{\rm TAMS}$ and rapidly expand into the stable RSG regime during core helium burning. In contrast, lower-metallicity stars have smaller $R_{\rm TAMS}$ and advance to more evolved core helium or carbon-burning stages while retaining compact envelopes, thereby preventing expansion into the RSG regime during core helium burning. Our results explain the origin of the critical metallicity and offer insight into the evolution of metal-poor massive stars in the early universe.
\end{abstract}


\section{Introduction}\label{sec:intro}
In the first paper of this series \citep[][hereafter Paper I]{ou2023}, we presented a grid of $\sim$2000 non-rotating massive star models and identified a critical metallicity at around $Z\sim 0.001$ for the formation of cool supergiants (i.e., red or yellow). We found that stars with metallicities below this threshold typically do not evolve into cool supergiants during the core-helium (He) burning phase. Our finding is consistent with various earlier stellar models, which likewise show that low-$Z$ stars tend to remain blue supergiants (BSGs) rather than evolving into red supergiants (RSGs) \citep{brunish1982,arnett1991,baraffe1991,brocato1993,hirschi2007,eleid2009,limongi2017,groh2019}, but we further identified a well-defined metallicity threshold for this transition. This bifurcation in evolutionary paths has a significant impact on stellar mass loss, as RSGs exhibit considerably higher mass-loss rates than main-sequence stars and BSGs. In this paper, we investigate the physical origin of this critical metallicity. 

An early attempt to explain the metallicity dependence of RSG formation was made by \citet{ritossa1996}, who argued that the envelopes of low-metallicity stars cannot effectively trap radiation energy from central burning because of reduced opacity, and therefore cannot evolve into RSGs.  This interpretation is based on a framework in which thermal imbalance within the envelope, governed by the core luminosity and the envelope opacity, drives the inflation of the stellar envelope to red-giant or supergiant dimensions \citep{Renzini1984,renzini1992,renzini1994,renzini2023}. However, this explanation for RSG formation based on heat absorption has been shown to be inconsistent with subsequent studies \citep{iben1993,sugimoto2000,faulkner2005,bertolami2022,ou2024}.

Understanding the metallicity effect on RSG formation is closely tied to the long-standing question: Why do stars become RSGs? Our recent study \citep{ou2024} proposed that envelope expansion toward the RSG phase follows the refined "mirror principle,"  under which there exist two pathways for stars to evolve into RSGs -- direct expansion during core contraction, and continued expansion after core contraction governed by nuclear energy generation rates. Building on those results, we now aim to investigate how metallicity influences RSG\footnote{We briefly clarify our use of the terms "red supergiant" and "cool supergiant." In Paper I, we adopted the term cool supergiant to describe massive stars that expand to $\gtrsim 1{,}000\, R_{\odot}$. We avoided the conventional term red supergiant because very massive stars (e.g., $\gtrsim 50\, M_{\odot}$) often maintain effective temperatures above $4{,}000\,{\rm K}$, placing them in the yellow rather than strictly defined red regime. Nevertheless, from a physical standpoint in terms of stellar radius, such stars do not differ intrinsically from classical red supergiants. Moreover, in this paper, our focus is mainly on stars with $\lesssim 30\, M_{\odot}$, which cool below $4{,}000\,{\rm K}$ and fulfill the criteria of red supergiants. Therefore, we adopt the more common term red supergiant throughout this work.} formation to identify the physical origin of the critical metallicity using stellar models.

In Section~\ref{sec:models}, we briefly describe the models employed in this study.  In Section~\ref{sec:experiments}, we present a series of numerical experiments to investigate the metallicity effect on supergiant evolution. Section~\ref{sec:radius} identifies a threshold radius at the end of main sequence for RSG formation and explores its dependence on mass. To explore the connection between metallicity and this threshold radius, Section~\ref{sec:metallicity} demonstrates how metallicity influences stellar radii. In Section~\ref{sec:core}, we offer a physical interpretation of the threshold radius and critical metallicity based on He core evolution. Finally, Section~\ref{sec:conclusions} discusses the broader implications and concludes the paper.

\section{Models}\label{sec:models}

We begin with the grid of massive star models from Paper I, hereafter Grid (a), which spans initial masses from $11$ to $60\, M_{\odot}$ and metallicities from $Z=1\times 10^{-5}$ to $2\times 10^{-2}$. The models were computed using the Modules for Experiments in Stellar Astrophysics  \citep[MESA;][]{paxton2011,paxton2013,paxton2015,paxton2018,paxton2019,jermyn2023} version No. 10108. Each model evolves from the pre-main-sequence stage using the "create\_zams" module with the “basic” nuclear reaction network, and then proceeds from the zero-age main sequence (ZAMS) using the “approx21 (Cr60)” network, which includes 21 isotopes. The models adopt Henyey’s mixing-length theory with a mixing-length parameter of $\alpha_{\rm MLT} = 1.5$. Semi-convection is included with an efficiency of $\alpha_{\rm sc} = 0.01$. Exponential overshooting is implemented following \citet{Herwig2000}, with $f_{\rm ov} = 0.001$ in non-burning and H-burning regions, $f_{\rm ov} = 0$ in He- and metal-burning regions, and $f_{0,\,\rm ov} = 0.0005$ in all regions. Further details of the model setup are provided in Section~2 of Paper~I.

Using the $25\,M_{\odot}$ models from Grid (a), we carry out a series of controlled experiments in which we individually modify several metallicity-dependent factors, such as opacity and nuclear reaction rates, to assess their impact on supergiant evolution. These numerical experiments are designed to separate the influence of specific physical parameters and are not intended to represent realistic stellar evolution.

To control opacity systematically, we vary the MESA parameter $Z_{\rm base}$, which specifies the base metallicity used to query the opacity tables. This parameter affects the opacity alone and is independent of the actual stellar metallicity. For example, a model with $Z = 0.02$ and $Z_{\rm base} = 0.001$ retains its true metallicity of 0.02 but evolves with opacities comparable to that of a star with $Z = 0.001$. This method helps to separate the effects of opacity from other metallicity-dependent influences, such as changes in nuclear energy generation and the mean molecular weight. To avoid confusion, we refer to $Z_{\rm base}$ as $\zeta_{\kappa}$ throughout this paper.

Nuclear reaction rates are modified using the \texttt{reaction\_for\_special\_factor} option in the \texttt{star\_job} section of the MESA inlist. For the CNO cycle, we adopt the parameter $\eta_{\rm CNO}$ to linearly scale the rates of five key reactions included in the approx21 nuclear reaction network:  (1) $^{12}{\rm C}+^1{\rm H} \rightarrow ^{13}{\rm N}+\gamma$, (2) $^{14}{\rm N}+^1{\rm H} \rightarrow ^{15}{\rm O}+\gamma$, (3) $^{16}{\rm O}+^1{\rm H} \rightarrow ^{17}{\rm F}+\gamma$, (4) $^{15}{\rm N}+^1{\rm H} \rightarrow ^{16}{\rm O}+\gamma$, and (5) $^{15}{\rm O}+^1{\rm H} \rightarrow ^{12}{\rm C}+^{4}{\rm He}$. In addition, we use another factor $\eta_{3\alpha}$ to linearly scale the reaction rate of the triple-alpha process.

We also perform an experiment in which the CNO reaction rates are modified only during the post-main-sequence shell H-burning phase. The adjustment is applied after the He core forms and is restricted to regions outside the core, which MESA defines as zones with $X_{\rm H} > 0.1$. In this experiment, we linearly scale the CNO reaction rates in the H burning shell by a factor $\eta_{\rm shellCNO}$.

In total, we construct five grids of MESA models covering different parameter spaces, comprising 6,896 models:

\begin{enumerate}
    \item[(a)] Initial stellar mass ($M_i$) and metallicity ($Z$), from Paper I. 
    \item[(b)] Metallicity ($Z$) and opacity (set by $\zeta_{\kappa}$). 
    \item[(c)] Hydrogen-burning rate (scaled by $\eta_{\rm CNO}$) and opacity (set by $\zeta_{\kappa}$). 
    \item[(d)] Shell Hydrogen-burning rate (scaled by $\eta_{\rm shellCNO}$ for the burning shell) and opacity (set by $\zeta_{\kappa}$).
    \item[(e)] Helium-burning rate (scaled by $\eta_{3\alpha}$) and opacity (set by $\zeta_{\kappa}$).

\end{enumerate}
Grids (b)--(e) serve as controlled experiments for separating the effects of specific stellar parameters. In each grid, only two parameters are varied, while all others are kept identical to Grid~(a). Grids (b)--(e) are based on $25\,M_{\odot}$ models, with Grids (c)--(e) further restricted to $Z = 0.001$. The full set of model parameters is summarized in Table~\ref{tab:model_grids}.

\begin{deluxetable*}{c|c|c|c|c|c|c}
\tabletypesize{\scriptsize}
\tablecaption{Summary of model grids and varied parameters\label{tab:model_grids}}
\tablewidth{0pt}
\setlength{\tabcolsep}{3pt} 

\tablehead{
\colhead{Grid} &
\colhead{$M_i\ (M_\odot)$} &
\colhead{$Z$} &
\colhead{$\zeta_{\kappa}$} &
\colhead{$\eta_{\rm CNO}$} &
\colhead{$\eta_{\rm shellCNO}$} &
\colhead{$\eta_{3\alpha}$}
}

\startdata
(a) &
\cellbox{1.8cm}{11, 12, 13, $\ldots$, 59, 60} &
\cellbox{2.5cm}{
$(1,\,2,\,\ldots,\,9)\times10^{-5}$\\
$(1,\,2,\,\ldots,\,9)\times10^{-4}$\\
$(1,\,2,\,\ldots,\,9)\times10^{-3}$\\
$(1.0,\,1.1,\,\ldots,\,2.0)\times10^{-2}$
} &
1.0 & 1.0 & 1.0 & 1.0 \\
\hline
(b) &
25 &
\cellbox{2.5cm}{
$(1,\,2,\,\ldots,\,9)\times10^{-5}$\\
$(1,\,2,\,\ldots,\,9)\times10^{-4}$\\
$(1,\,2,\,\ldots,\,9)\times10^{-3}$\\
$(1.0,\,1.1,\,\ldots,\,2.0)\times10^{-2}$
} &
\cellbox{2.5cm}{
$(1,\,2,\,\ldots,\,9)\times10^{-5}$\\
$(1,\,2,\,\ldots,\,9)\times10^{-4}$\\
$(1,\,2,\,\ldots,\,9)\times10^{-3}$\\
$(1.0,\,1.1,\,\ldots,\,2.0)\times10^{-2}$
} &
1.0 & 1.0 & 1.0 \\
\hline
(c) &
25 &
0.001 &
\cellbox{2.5cm}{
$(1,\,2,\,\ldots,\,9)\times10^{-5}$\\
$(1,\,2,\,\ldots,\,9)\times10^{-4}$\\
$(1,\,2,\,\ldots,\,9)\times10^{-3}$\\
$(1,\,2,\,3,\,4,\,5)\times10^{-2}$
} &
\cellbox{2.5cm}{
$(1,\,2,\,\ldots,\,9)\times10^{-3}$\\
$(1,\,2,\,\ldots,\,9)\times10^{-2}$\\
$(1,\,2,\,\ldots,\,9)\times10^{-1}$\\
$(1,\,2,\,\ldots,\,10)$
} &
1.0 & 1.0 \\
\hline
(d) &
25 &
0.001 &
\cellbox{2.5cm}{
$(1,\,2,\,\ldots,\,9)\times10^{-5}$\\
$(1,\,2,\,\ldots,\,9)\times10^{-4}$\\
$(1,\,2,\,\ldots,\,9)\times10^{-3}$\\
$(1,\,2,\,3,\,4,\,5)\times10^{-2}$
} &
1.0 &
\cellbox{2.5cm}{
$(1,\,2,\,\ldots,\,9)\times10^{-3}$\\
$(1,\,2,\,\ldots,\,9)\times10^{-2}$\\
$(1,\,2,\,\ldots,\,9)\times10^{-1}$\\
$(1,\,2,\,\ldots,\,10)$
} &
1.0 \\
\hline
(e) &
25 &
0.001 &
\cellbox{2.5cm}{
$(1,\,2,\,\ldots,\,9)\times10^{-5}$\\
$(1,\,2,\,\ldots,\,9)\times10^{-4}$\\
$(1,\,2,\,\ldots,\,9)\times10^{-3}$\\
$(1,\,2,\,3,\,4,\,5)\times10^{-2}$
} &
1.0 & 1.0 &
\cellbox{2.5cm}{
$(1,\,2,\,\ldots,\,9)\times10^{-3}$\\
$(1,\,2,\,\ldots,\,9)\times10^{-2}$\\
$(1,\,2,\,\ldots,\,9)\times10^{-1}$\\
$(1,\,2,\,\ldots,\,10)$
} \\
\enddata

\tablecomments{
$M_i$: Initial stellar mass.\\
$Z$: Stellar metallicity.\\
$\zeta_{\kappa}$: Equivalent to the base metallicity ($Z_{\rm base}$) adopted for interpolating opacity in the tables, and should not be confused with the stellar metallicity $Z$.\\
$\eta_{\rm CNO}$: Scaling factor for the CNO cycle reaction rate.\\
$\eta_{\rm shellCNO}$: Scaling factor for the CNO cycle reaction rate, applied only to the H-burning shell.\\
$\eta_{3\alpha}$: Scaling factor for the triple-alpha reaction rate.
}
\end{deluxetable*}

\section{Numerical Experiments on Opacity and Nuclear Reaction Rates}\label{sec:experiments}

Metallicity influences stellar evolution through its effects on opacity, nuclear burning rates, and the mean molecular weight \citep[e.g.,][]{eleid2009}. In this section, we use Grids (b)--(e) to separate the roles of opacity and nuclear reaction rates in determining whether a star evolves toward the RSG phase. For each model in these grids, we evaluate its evolutionary outcome by computing the maximum radius attained during core He burning—before the onset of core carbon (C) burning—which we denote as $R_{\rm max}$.
The values of $R_{\rm max}$ are plotted in grid plots, from which we observe the division between models that reach RSG dimensions ($\log(R/R_{\odot}) \gtrsim 3$) and those that remain in the BSG regime ($\log(R/R_{\odot}) \lesssim 2$). We examine whether the boundary between these two regimes aligns along any threshold of opacity or nuclear reaction rate parameters.

\begin{figure}[tbh]
\centering
\includegraphics[width=\columnwidth]{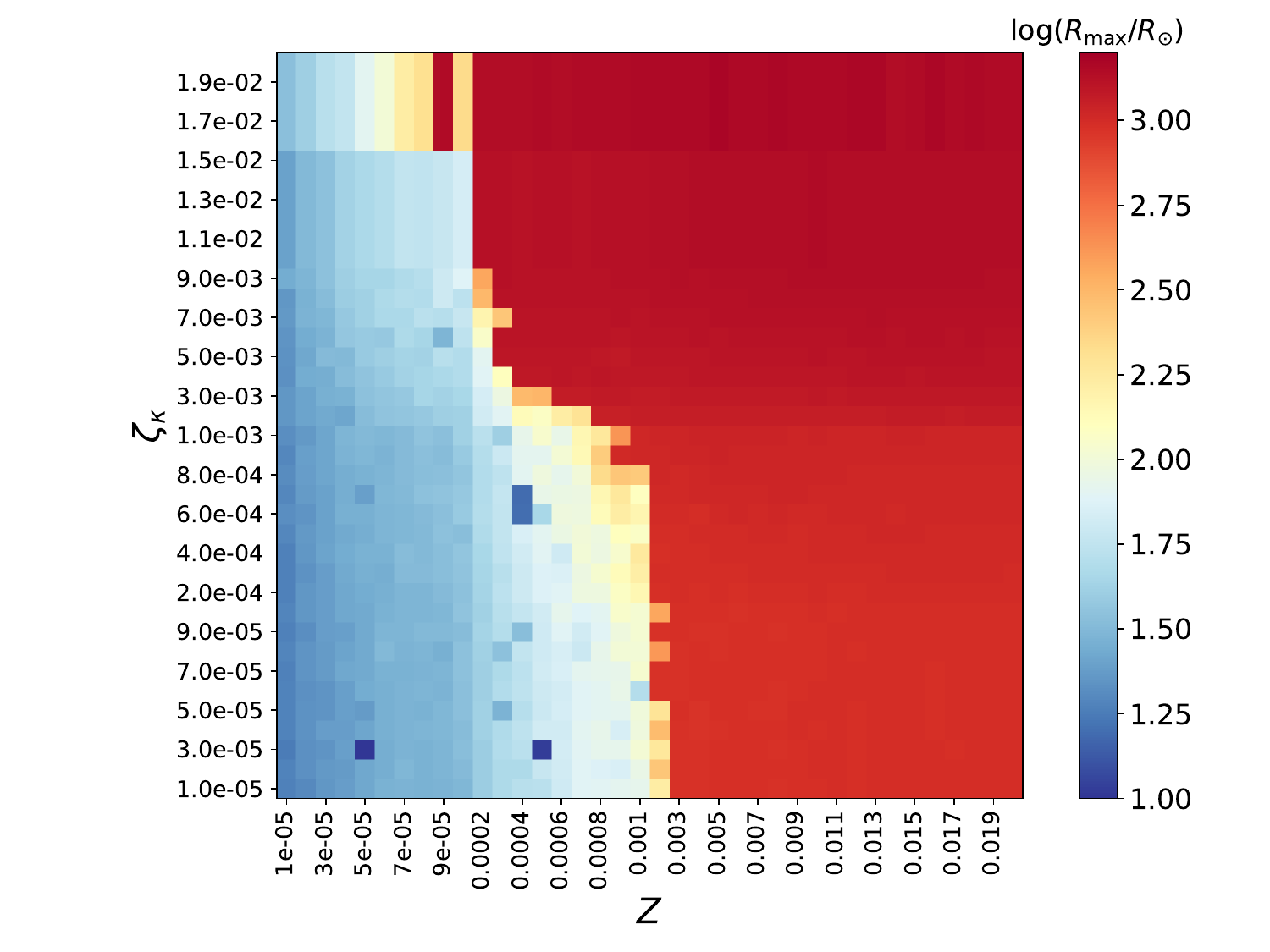}
\caption{Maximum stellar radius during the core He-burning stage ($R_{\rm max}$) for Grid (b), which includes models with an initial mass of $25\,M_{\odot}$, varying in metallicity ($Z$) and opacity parameter ($\zeta_{\kappa}$). The boundary between red and blue supergiants depends on $\zeta_{\kappa}$ in the range $\zeta_{\kappa} \gtrsim 10^{-3}$, but becomes nearly insensitive to $\zeta_{\kappa}$ when $\zeta_{\kappa} \lesssim 10^{-3}$.}
\label{fig:z-zbase}
\end{figure}
\begin{figure}[tbh]
\centering
\includegraphics[width=\columnwidth]{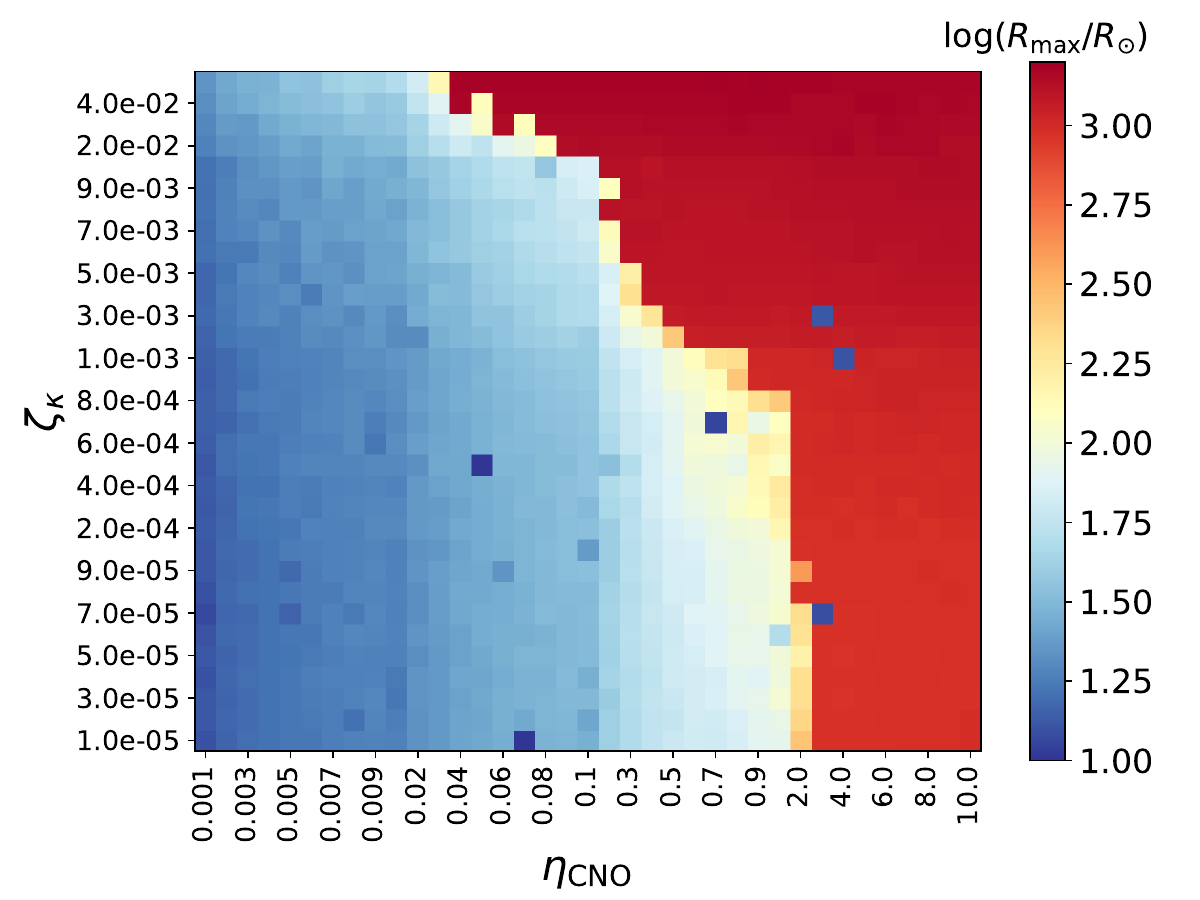}
\caption{Maximum stellar radius during the core He-burning stage ($R_{\rm max}$) for Grid (c): models with an initial mass of $25\,M_{\odot}$ and metallicity $Z=0.001$, computed for varying values of the CNO-cycle reaction rate (scaled by $\eta_{\rm CNO}$) and opacity parameter ($\zeta_{\kappa}$). Both $\eta_{\rm CNO}$ and $\zeta_{\kappa}$ influence the evolutionary outcome, determining whether the star becomes a red or blue supergiant.}
\label{fig:CNO-zbase}
\end{figure}
\begin{figure}[tbh]
\centering
\includegraphics[width=\columnwidth]{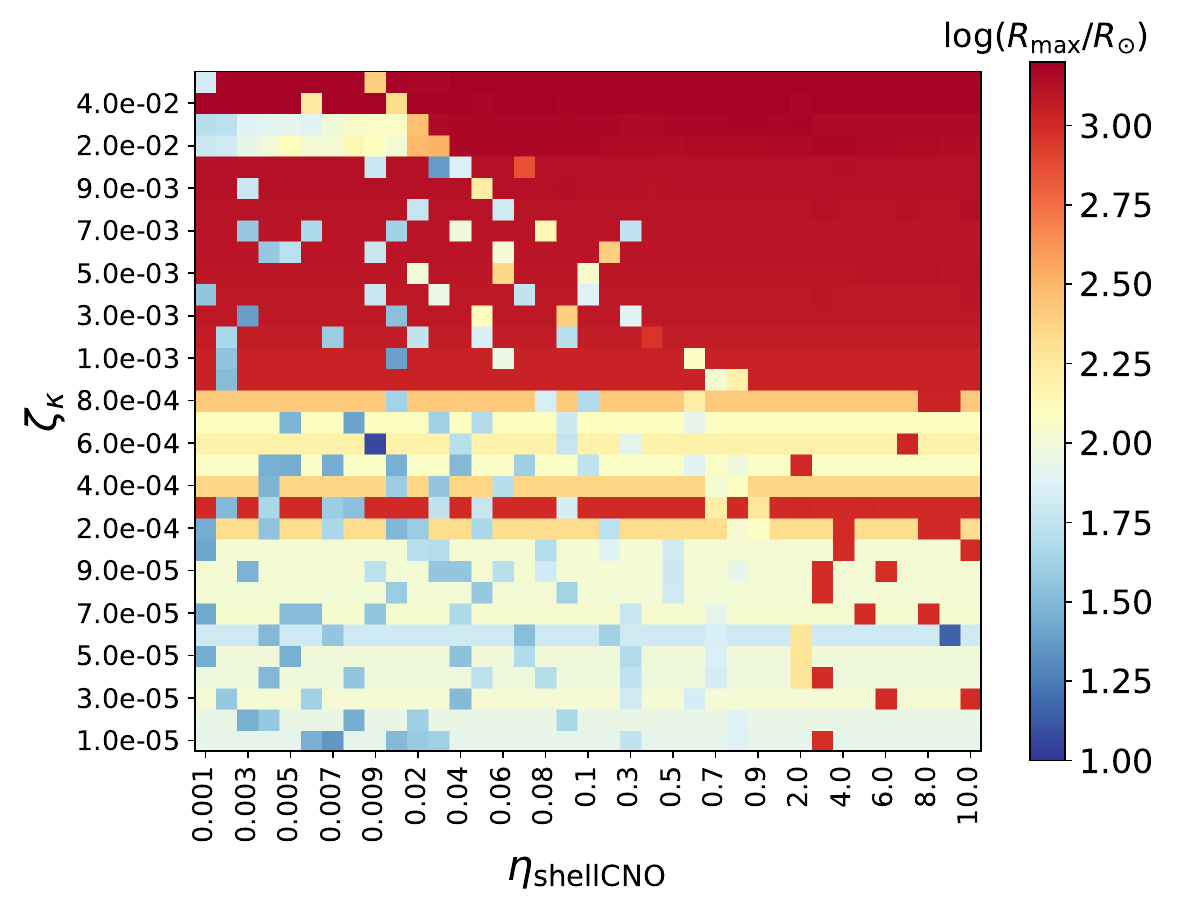}
\caption{Maximum stellar radius during the core He-burning stage ($R_{\rm max}$) for Grid (d): models with an initial mass of $25\,M_{\odot}$ and metallicity $Z=0.001$, computed for varying values of the CNO-cycle reaction rate during the shell H-burning stage (scaled by $\eta_{\rm shellCNO}$) and opacity parameter ($\zeta_{\kappa}$). For comparison, the parameter $\eta_{\rm CNO}$ used in Figure~\ref{fig:CNO-zbase} is applied throughout all evolutionary stages, whereas $\eta_{\rm shellCNO}$ in this figure is applied only during the shell H-burning stage. While $R_{\rm max}$ exhibits some scatter, the overall trend indicates that $\eta_{\rm shellCNO}$ has little impact on the supergiant outcome.}
\label{fig:HshellCNO-zbase}
\end{figure}
\begin{figure}[tbh]
\centering
\includegraphics[width=\columnwidth]{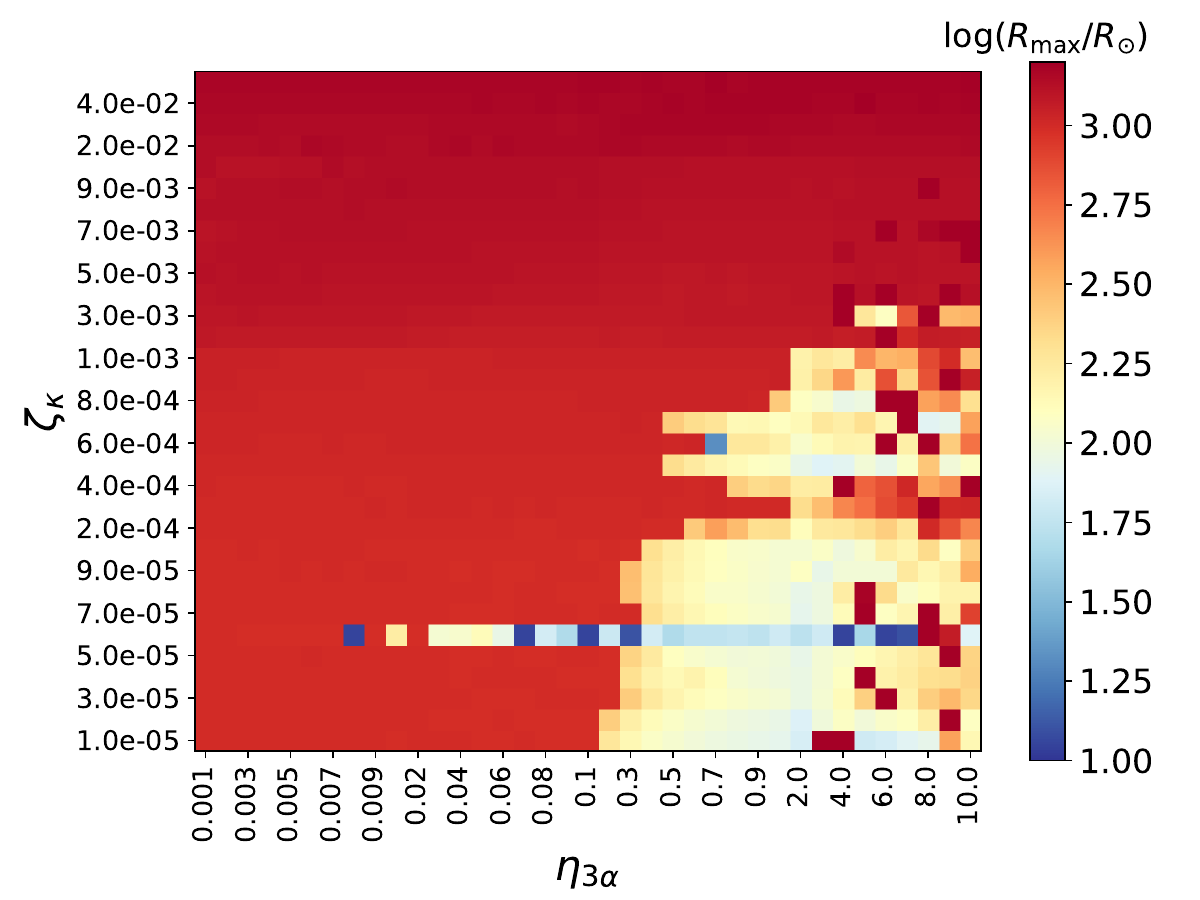}
\caption{Maximum stellar radius during the core He-burning stage ($R_{\rm max}$) for Grid (e): models with an initial mass of $25\,M_{\odot}$ and metallicity $Z=0.001$, computed for varying values of the triple-alpha reaction rate (scaled by $\eta_{3\alpha}$) and opacity parameter ($\zeta_{\kappa}$). Increasing $\eta_{3\alpha}$ does not promote evolution toward the RSG phase and may even suppress it in the low-opacity regime ($\zeta_{\kappa} \lesssim 10^{-3}$).}
\label{fig:talpha-zbase}
\end{figure}

\subsection{Grid (b): Metallicity and Opacity}
 
Grid (b) varies the metallicity and opacity for $25\,M_{\odot}$ models to test how opacity affects the supergiant outcome. The resulting values of $R_{\rm max}$ are shown in Figure~\ref{fig:z-zbase}, which reveals a clear separation between RSG and BSG outcomes (red and blue, respectively). When $\zeta_{\kappa}\gtrsim 10^{-3}$, opacity can shift the metallicity threshold for RSG formation, with higher opacities allowing RSGs to form at lower $Z$. However, for $\zeta_{\kappa} \lesssim 10^{-3}$, the boundary between RSG and BSG models becomes vertical, indicating that opacity has little influence on the evolutionary outcome in this regime. Notably, even with extremely low $\zeta_{\kappa}$ (a few times $10^{-5}$), stars with $Z \gtrsim 0.003$ still expand to RSG dimensions.
This implies that opacity may not be the only metallicity-dependent factor that determines RSG formation.

\subsection{Grid (c): Hydrogen Burning Rate and Opacity}

Metallicity also strongly affects nuclear burning. Using Grid (c), we examine the combined influence of two metallicity-dependent factors—opacity (set by $\zeta_{\kappa}$) and the hydrogen-burning rate (scaled by $\eta_{\rm CNO}$ and applied to both core H burning on the main sequence and shell H burning thereafter). As shown in Figure~\ref{fig:CNO-zbase}, the boundary between the red (RSG) and blue (BSG) regions indicates that both higher opacity and higher H burning rate promote the transition to RSGs. Even when the opacity is low (e.g., $\zeta_{\kappa}$ of a few $\times 10^{-5}$), the star can still evolve into an RSG if H burning is sufficiently efficient (i.e., $\eta_{\rm CNO} \geq 3$).

\subsection{Grid (d): Shell Hydrogen Burning Rate and Opacity}

The reaction rates of H burning can strongly influence whether a star evolves into an RSG, but it is unclear whether this sensitivity originates mainly from core or shell H burning. One might intuitively expect shell H burning in the post-main-sequence phase to drive envelope expansion. To test this, we construct Grid (d), where the CNO reaction rates are modified only during post-main-sequence shell H burning (scaled by $\eta_{\rm shellCNO}$) while leaving main-sequence burning unchanged.

Figure~\ref{fig:HshellCNO-zbase} presents the resulting $R_{\rm max}$ values. Despite some scatter, $\eta_{\rm shellCNO}$ has only a minor effect on the maximum radii of supergiants. This shows that the strong dependence of $R_{\rm max}$ on $\eta_{\rm CNO}$ seen in Grid (c) may not originate from shell H burning, but rather from modifications to core H burning during the main sequence. Thus, we obtain the intriguing result that nuclear burning in the main sequence can significantly influence the degree of envelope expansion later in evolution.

\subsection{Grid (e): Helium Burning Rate and Opacity}

In \citet{ou2024}, we showed that enhanced He burning can actually suppress the expansion toward the RSG phase. To further demonstrate this effect, we construct Grid (e), in which the triple-alpha reaction rate is scaled by a linear coefficient $\eta_{3\alpha}$, as shown in Figure~\ref{fig:talpha-zbase}.  In the lower-right region of the diagram, some models with low $\zeta_{\kappa}$ and high $\eta_{3\alpha}$ fail to evolve into RSGs. An increase in $\eta_{3\alpha}$ not only does not promote RSG formation but can even inhibit it, particularly at low opacity.

In summary, we conducted a series of experiments and find that both higher opacity and a stronger core H-burning rate during the main sequence favor expansion into the RSG regime, whereas enhanced core He burning suppresses it. These trends explain why higher-metallicity stars more readily become RSGs. However, although opacity and core H burning jointly shape the outcome, the common underlying trigger that enables the RSG transition remains unclear, which motivates further investigation in the next section.

\begin{figure*}[tbh]
\centering
\includegraphics[scale=0.5]{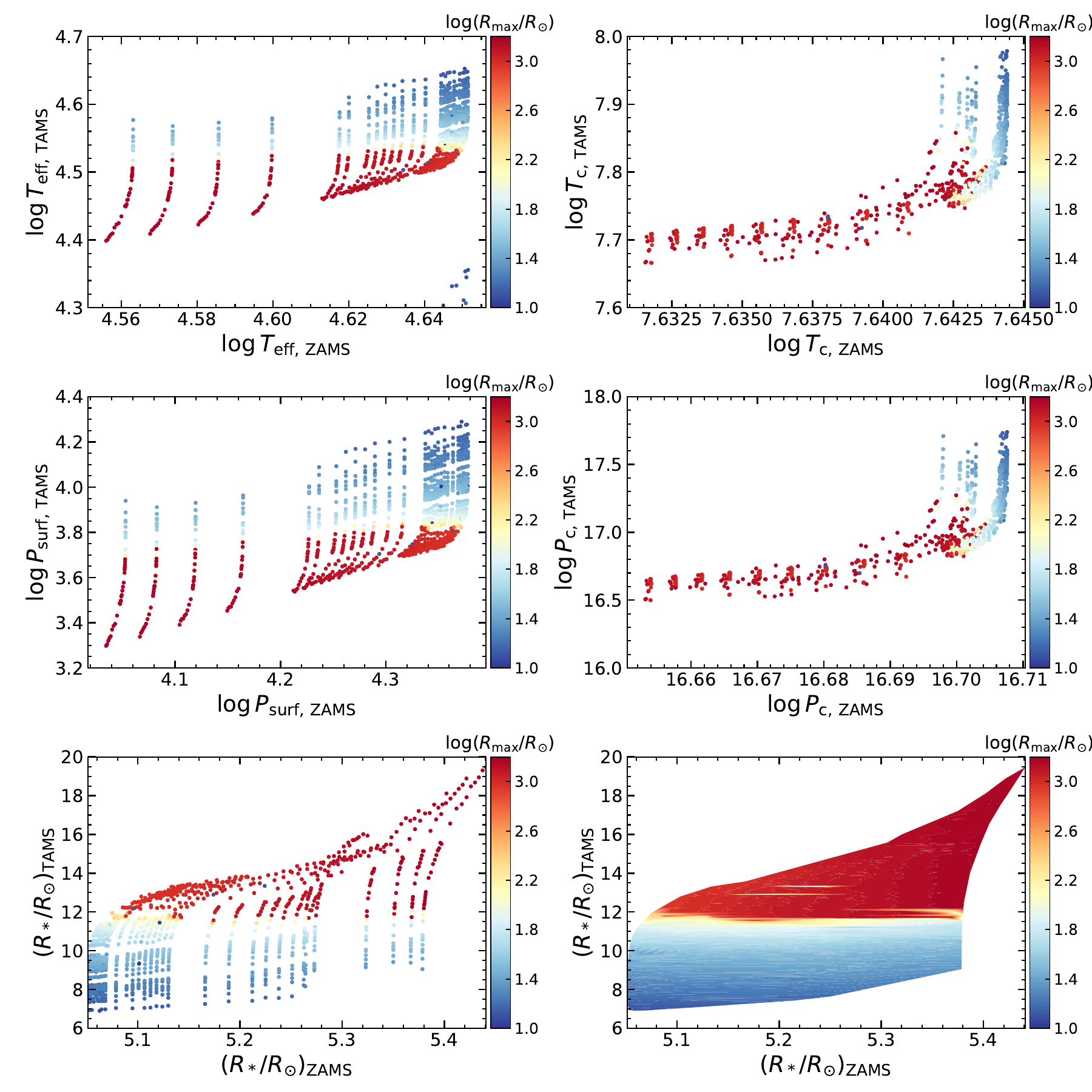}
\caption{Effective temperature ($T_{\rm eff}$), central temperature ($T_{\rm c}$), surface pressure ($P_{\rm surf}$), central pressure ($P_{\rm c}$), and stellar radius ($R_*$) at the ZAMS and TAMS for $25\,M_{\odot}$ models from Grid (c). In each panel, the ZAMS value is shown on the X-axis and the TAMS value on the Y-axis. Points are colored by the maximum stellar radius ($R_{\rm max}$) reached during core He burning, with red indicating RSG-scale radii and blue indicating BSG-scale radii. The bottom-right panel shows a smoothed contour representation of the same $R_*$ data displayed in the bottom-left panel. Notably, models with TAMS radii $\gtrsim 12\, R_{\odot}$ predominantly evolve into RSGs, whereas those below this threshold remain BSGs.}
\label{fig:cno_zbase_radius}
\end{figure*}

\section{Threshold for RSG Formation}\label{sec:radius}

We now aim to develop a unified picture of how opacity and core H-burning rates jointly modify the stellar structure and therefore determine whether a star evolves into an RSG. Specifically, we seek to identify the key structural condition established during the main-sequence phase that ultimately enables the formation of RSG.

To this end, we examine stellar parameters at two evolutionary points—the ZAMS and the terminal-age main sequence (TAMS)—and assess how they correlate with the maximum radius ($R_{\rm max}$) reached during core He burning. In our MESA models, the ZAMS is defined using the "create\_zams" template, which generates a ZAMS model that serves as the initial condition for subsequent evolution. Accordingly, we extract the ZAMS parameters from the first timestep of each evolutionary track. The TAMS is defined as the last timestep before the emergence of a He core, identified by the onset of a non-zero He-core mass in the MESA output.

Figure~\ref{fig:cno_zbase_radius} uses Grid (c), in which the initial mass is fixed at $25\,M_{\odot}$, to show how ZAMS and TAMS stellar properties correlate with the supergiant outcome. Each panel displays one stellar parameter—effective temperature ($T_{\rm eff}$), central temperature ($T_{\rm c}$), surface pressure ($P_{\rm surf}$), central pressure ($P_{\rm c}$), or stellar radius ($R_*$)—with its ZAMS value on the X-axis and its TAMS value on the Y-axis. Each model in Grid (c) is plotted on these panels and colored by its resulting $R_{\rm max}$.

Our aim is to identify the boundary separating models that evolve into RSGs (red points) from those that remain BSGs (blue points). The results show that the red–blue division is well defined in the surface parameters ($T_{\rm eff}$, $P_{\rm surf}$, and $R_*$), but not in the core parameters ($T_{\rm c}$ and $P_{\rm c}$). Moreover, these division lines appear nearly horizontal, aligning with fixed TAMS surface values rather than ZAMS values. In particular, the TAMS radius emerges as a strong indicator of RSG formation: stars with any ZAMS radius can become RSGs, but only those whose TAMS radius exceeds a threshold of $\sim 12\,R_{\odot}$ ultimately evolve into RSGs.

\begin{figure}[tbh]
\centering
\includegraphics[scale=0.5]{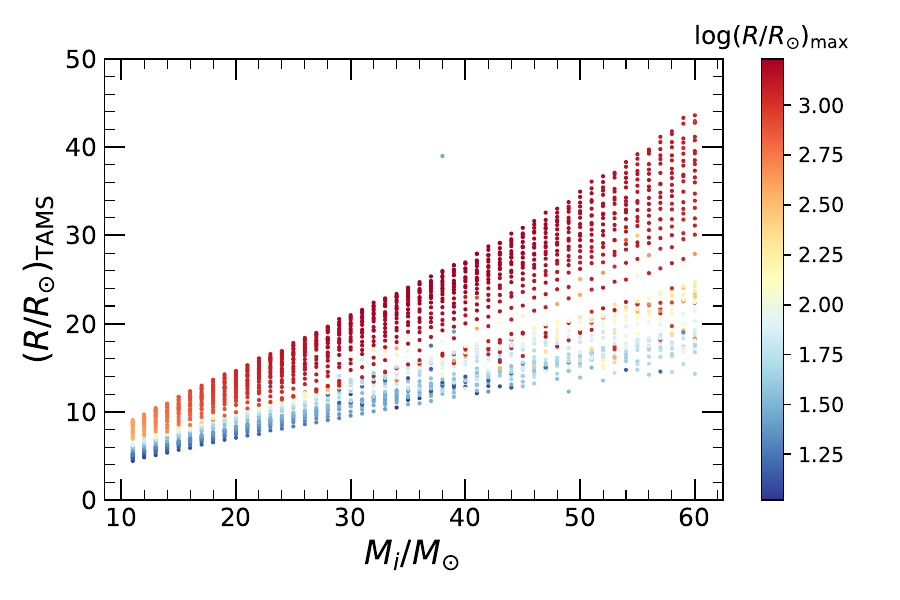}
\includegraphics[scale=0.5]{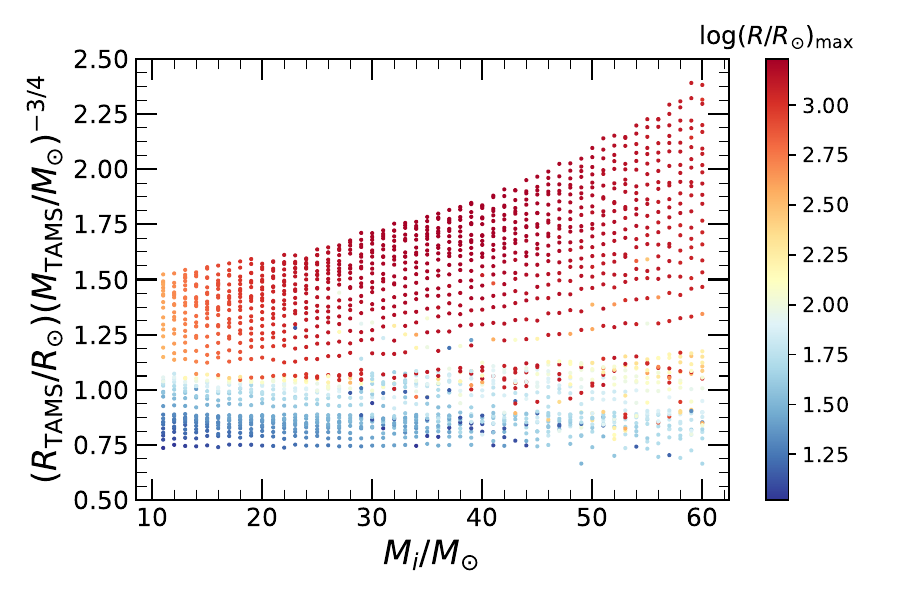}
\caption{The relationship between the TAMS radius and initial stellar mass ($M_{\rm i}$), with the colors of the data points representing $R_{\rm max}$.}
\label{fig:m_radius_TAMS}
\end{figure}

Figure~\ref{fig:m_radius_TAMS} uses Grid (a) to conduct a similar analysis for stars with different initial masses. The red-blue boundary is no longer located at a constant TAMS radius but increases systematically with stellar mass. As shown in the right panel, we obtain an approximate criterion for RSG formation:

\begin{equation}
\bigg(\frac{R_{\rm TAMS}}{R_{\odot}}\bigg)\gtrsim \bigg(\frac{M_{\rm TAMS}}{M_{\odot}}\bigg) ^{3/4} ,
\end{equation}

\noindent where $R_{\rm TAMS}$ and $M_{\rm TAMS}$ are the radius and the mass, respectively, at the TAMS. Note that $M_{\rm TAMS}$ differs only slightly from the initial mass ($M_{\rm i}$), as mass loss during the main sequence is typically modest.

Therefore, we arrive at a unified picture of how opacity and the core H-burning rate promote RSG formation: both factors modify the stellar structure during the main-sequence phase, leading to different TAMS radii, which in turn serve as the decisive quantity for RSG formation. Regardless of which process alters the TAMS radius, a star can evolve into an RSG as long as its TAMS radius exceeds a critical threshold. However, the physical mechanism linking metallicity, the TAMS radius, and RSG formation remains unclear. In the following sections, we address two key questions:\\
(1) How does metallicity affect the TAMS radius?\\
(2) How does the TAMS radius influence post-main-sequence expansion and define a threshold for RSG formation?\\
We begin by quantifying how metallicity affects stellar radius in Section~\ref{sec:metallicity}, and then analyze how the TAMS radius regulates supergiant expansion in Section~\ref{sec:core}.

\section{Metallicity Effect on Stellar Radius}\label{sec:metallicity}

This section addresses this question: How does metallicity affect the TAMS radius? We begin by analyzing evolutionary tracks, and then apply homology relations to perform a dimensional analysis.

\subsection{Metallicity Effect on Stellar Evolutionary Tracks}

\begin{figure}[tbh]
\centering
\includegraphics[width=\columnwidth]{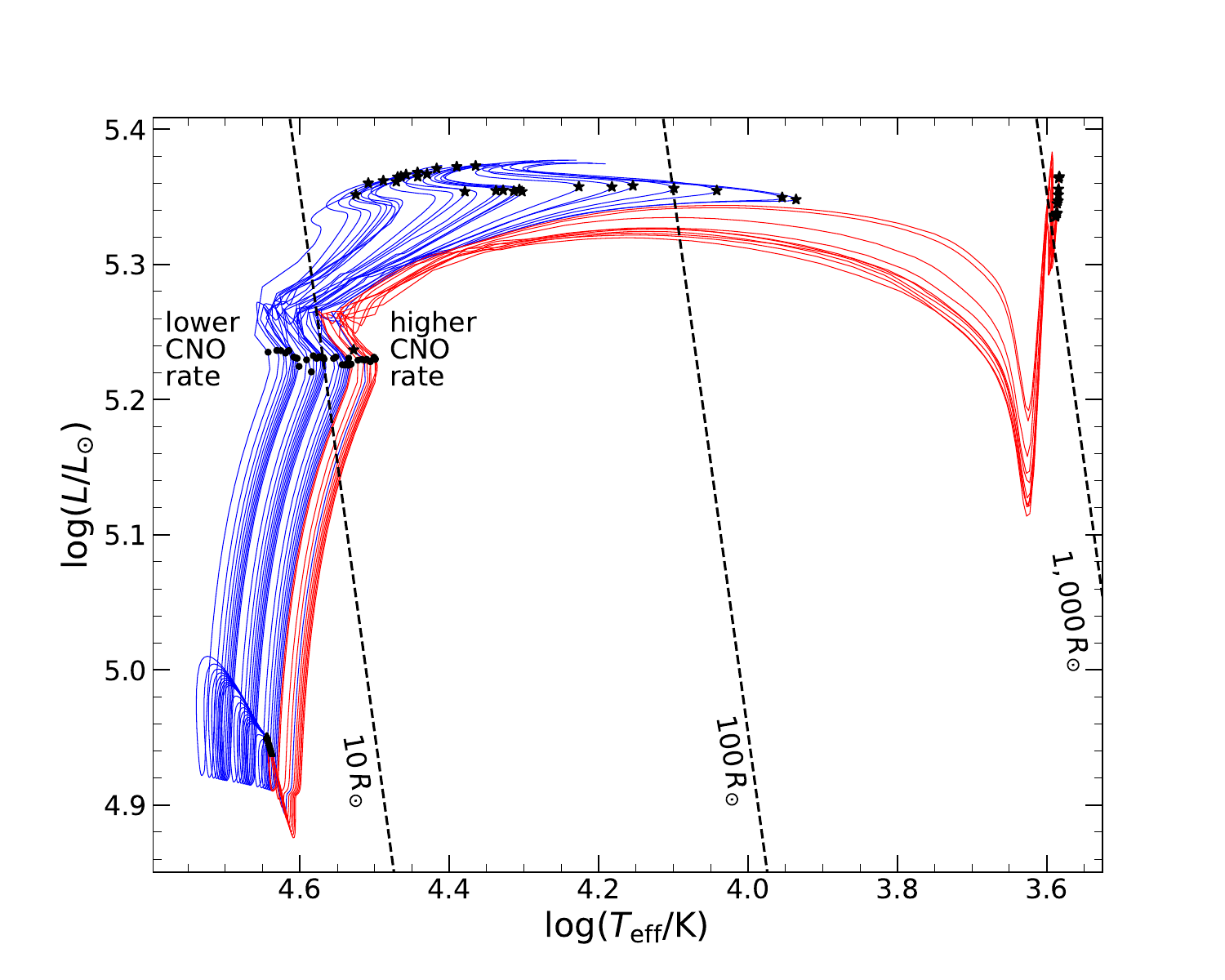}
\caption{Evolutionary tracks of $25\,M_{\odot}$, $Z=0.001$ stars with different $\eta_{\rm CNO}$ (subset of Grid (c)) on the HR diagram. Dashed lines indicate constant-radius contours at 10, 100, and $1000\,R_{\odot}$. Black triangles mark the ZAMS, black circles indicate the TAMS, and black stars denote the maximum stellar radius reached during the core He-burning phase. Tracks that evolve into RSGs are shown in red, while those that remain BSGs are shown in blue.}
\label{fig:HR_Zbase_multicno}
\end{figure}

\begin{figure}[tbh]
\centering
\includegraphics[width=\columnwidth]{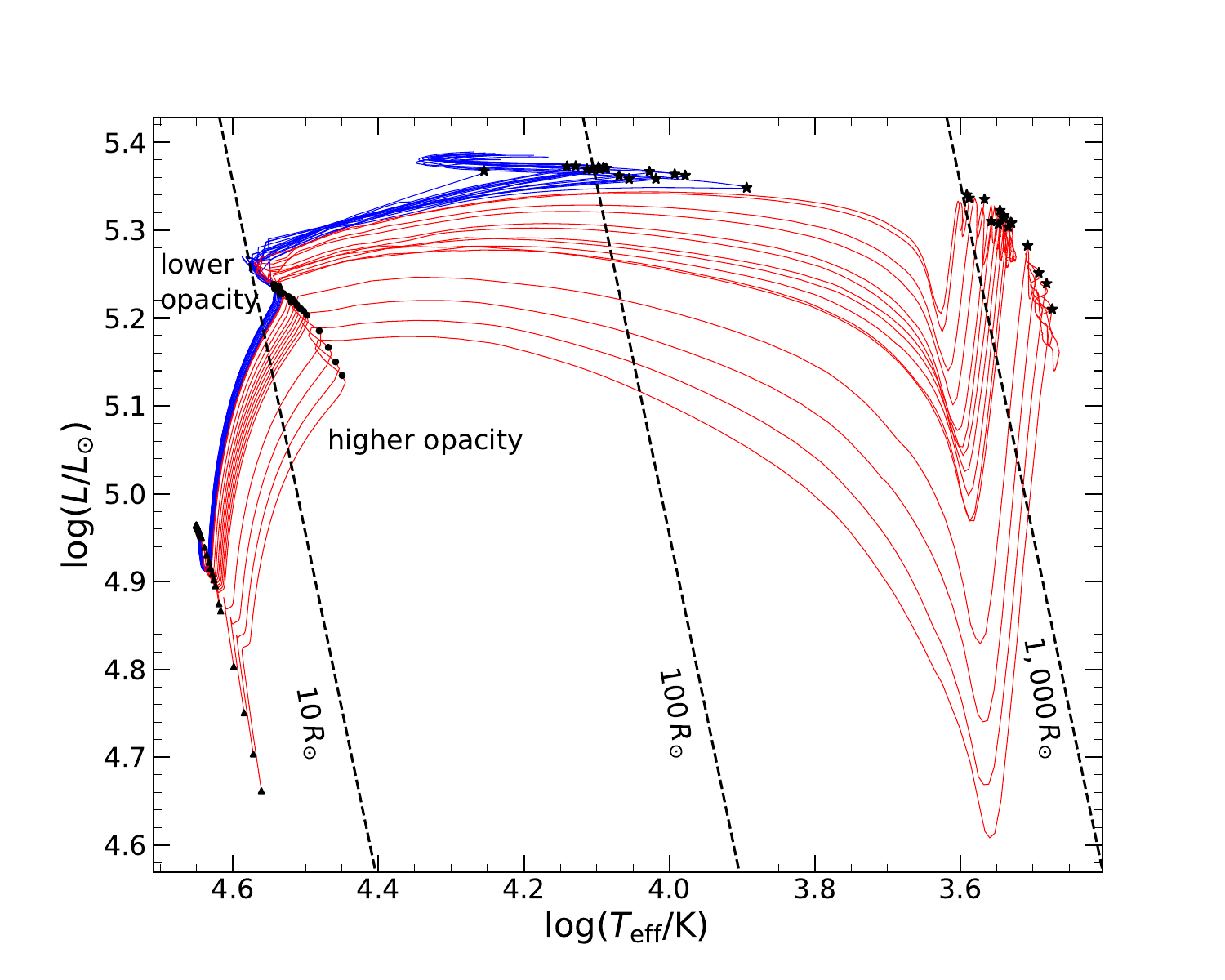}
\caption{Evolutionary tracks of $25\,M_{\odot}$, $Z=0.001$ stars with different $\zeta_{\kappa}$ (subset of Grid (c)) on the HR diagram. Dashed lines indicate constant-radius contours at 10, 100, and $1000\,R_{\odot}$. Black triangles mark the ZAMS, black circles indicate the TAMS, and black stars denote the maximum stellar radius reached during the core He-burning phase. Tracks that evolve into RSGs are shown in red, while those that remain BSGs are shown in blue.}
\label{fig:HR_cno_multiZbase}
\end{figure}

We have shown that $\eta_{\rm CNO}$ and $\zeta_{\kappa}$—both metallicity-dependent factors—influence whether a star evolves into an RSG by modifying its $R_{\rm TAMS}$. To examine how each parameter affects $R_{\rm TAMS}$, we present the evolutionary tracks for two subsets of the Grid (c) models: one in which $\eta_{\rm CNO}$ varies while $\zeta_{\kappa}$ is fixed (Figure~\ref{fig:HR_Zbase_multicno}), and another in which $\zeta_{\kappa}$ varies while $\eta_{\rm CNO}$ is held constant (Figure~\ref{fig:HR_cno_multiZbase}).

In Figure~\ref{fig:HR_Zbase_multicno}, the ZAMS positions are not significantly affected by changes in $\eta_{\rm CNO}$, but the subsequent evolutionary tracks exhibit a systematic shift in $T_{\rm eff}$ across different values of $\eta_{\rm CNO}$. Throughout the main-sequence phase and up to the TAMS, higher values of $\eta_{\rm CNO}$ result in lower $T_{\rm eff}$, while the luminosities remain similar. Consequently, stars with enhanced H-burning rates attain larger $R_{\rm TAMS}$.

In Figure~\ref{fig:HR_cno_multiZbase}, variations in $\zeta_{\kappa}$ produce noticeable differences in luminosity from the ZAMS onward, accompanied by moderate shifts in $T_{\rm eff}$. In general, a higher opacity (larger $\zeta_{\kappa}$) leads to a lower luminosity. The ZAMS locations lie nearly parallel to iso-radius lines, indicating that the stellar radius is only weakly sensitive to opacity at this stage. As models evolve, the higher opacity gradually decreases $T_{\rm eff}$. By the TAMS—just before expansion into the supergiant phase—models with higher opacity exhibit systematically lower $T_{\rm eff}$ and larger radii.

\begin{figure}[tbh]
\centering
\includegraphics[width=\columnwidth]{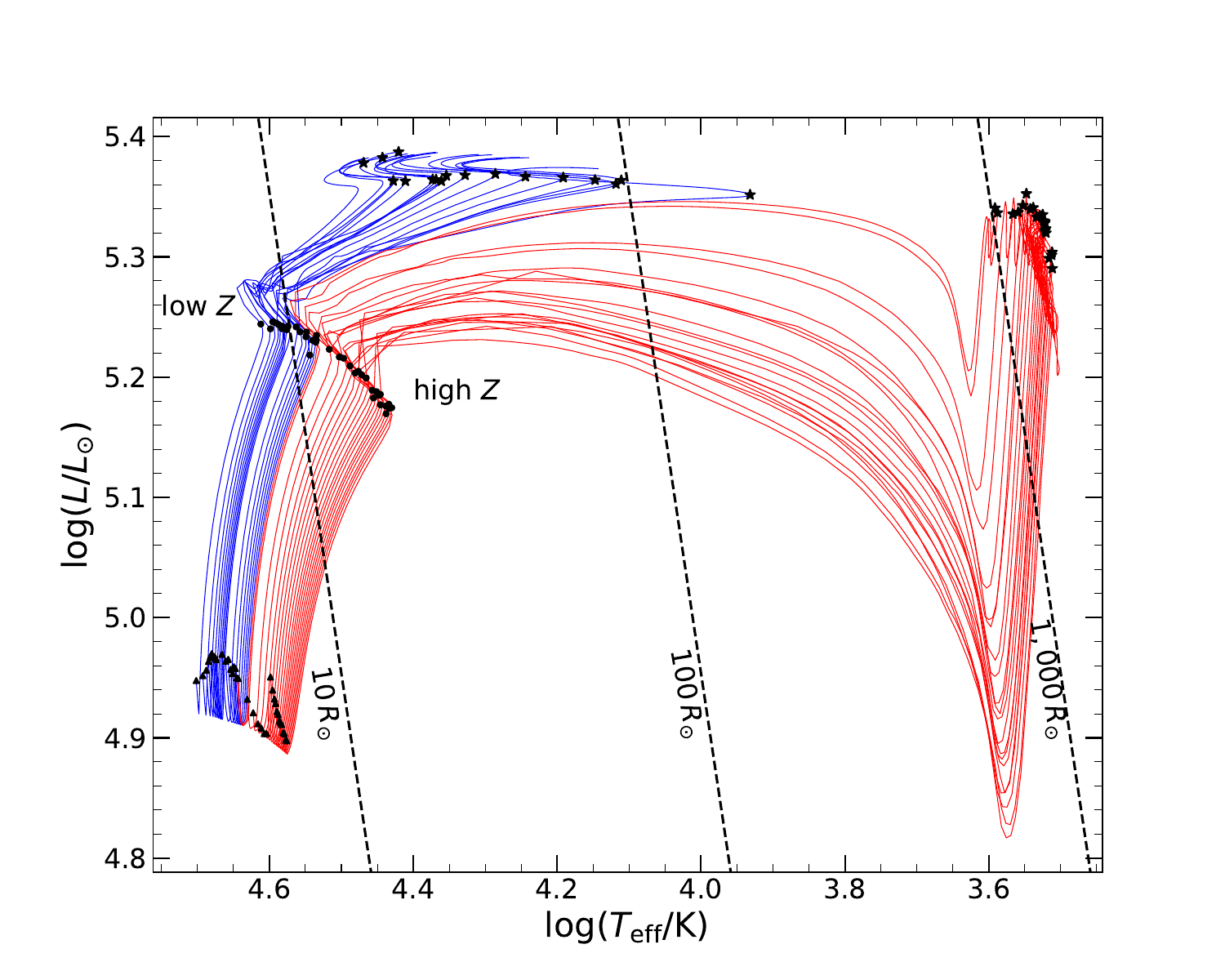}
\caption{Evolutionary tracks of $25\,M_{\odot}$ stars with different metallicities (subset of Grid (a)) on the HR diagram. Dashed lines indicate constant-radius contours at 10, 100, and $1000\,R_{\odot}$. Black triangles mark the ZAMS, black circles indicate the TAMS, and black stars denote the maximum stellar radius reached during the core He-burning phase. Tracks that evolve into RSGs are shown in red, while those that remain BSGs are shown in blue.}
\label{fig:HR}
\end{figure}

These results show that during the main-sequence phase, both increased opacity and a higher H-burning rate shift stellar evolutionary tracks toward cooler effective temperatures and larger radii on the HR diagram. At higher metallicities—where both opacity and the H-burning rate are enhanced—stars are therefore expected to develop systematically larger $R_{\rm TAMS}$. This cumulative metallicity effect is illustrated in Figure~\ref{fig:HR}, which presents evolutionary tracks of $25\,M_{\odot}$ models with different metallicities from Grid (a). Although the ZAMS positions display a relatively irregular distribution with $Z$, the tracks evolve towards a clearer metallicity trend near the TAMS: higher-$Z$ models reach lower $T_{\rm eff}$ at the TAMS, producing systematically larger $R_{\rm TAMS}$. This behavior reflects the influence of metallicity through its combined effects on opacity and nuclear burning.


\subsection{Homology Relation between Metallicity and Radius}

The effects of opacity and nuclear energy generation rates on stellar radius can be understood through homology relations \citep[see, e.g.,][]{maeder2009,kippenhahn2012}. The central idea of the homology method is to compare two stars with similar internal structures. At a given relative mass coordinate, defined as $\xi = m/M$, where $M$ is the total stellar mass and $m$ is the enclosure mass, the fundamental physical quantities in the two stars can be transformed with simple scaling constants. In the analysis below, we simply use the global stellar parameters such as $M$ or the average values within the entire star to express the homology relations to avoid complexity.

The four differential equations governing the stellar structure can be transformed into four homology relations. The mass conservation equation yields 
\begin{equation}
    M \sim \rho R^3,
\end{equation}
where $M$ is the total stellar mass, $\rho$ is the mean density, and $R$ is the stellar radius. The hydrostatic equilibrium equation gives
\begin{equation}
    P \sim M^2 R^{-4},
\end{equation}
where $P$ is the central pressure. The energy conservation leads to 
\begin{equation}
    L \sim \epsilon M,
\end{equation}
where $L$ is the luminosity and $\epsilon$ is the average energy generation rate per unit mass. The radiative energy transport equation is:
\begin{equation}
    \frac{dT}{dm} = -\frac{3}{64\pi^2 ac}\frac{\kappa l}{r^4T^3},
\end{equation}
where $T$ is the temperature, $\kappa$ is the opacity, $l$ is the local luminosity, $a$ is the radiation density constant, and $c$ is the speed of light. This equation results in the fourth homology relation
\begin{equation}
    R \sim (\kappa L M)^{1/4} T^{-1}.
\end{equation}
Combining Equations (4) and (6), the two relations associated with energy, we obtain 
\begin{equation}
RT \sim (\kappa \epsilon)^{1/4} M^{1/2}.
\end{equation}
Thus, at fixed stellar mass, increasing either $\epsilon$ or $\kappa$ leads to a larger product $R T$. Since both $\epsilon$ and $\kappa$ depend on the metallicity $Z$, variations in $Z$ can significantly affect the stellar structure.

To derive an explicit expression of $R$ as a function of $Z$, we adopt additional assumptions for the gas properties in a star. Following \citet{maeder2009}, we assume:

\begin{itemize}
    \item An ideal gas equation of state: $\rho = \frac{\mu m_{\rm H}}{k} P T^{-1}$, where $\mu$ is the mean molecular weight, $m_{\rm H}$ is the mass of a H atom, and $k$ is the Boltzmann constant.  
    \item Kramers' opacity law: $\kappa = \kappa_0 \rho T^{-3.5}$, where $\kappa_0$ is a coefficient that depends on its composition;
    \item A power-law form for nuclear energy generation rate: $\epsilon = \epsilon_0 \rho T^{\nu}$, where $\epsilon_0$ is a coefficient that depends on its composition, and the exponent $\nu$ depends on the dominant nuclear reaction. For the CNO cycle, $\nu = 17$.
\end{itemize}

\begin{figure}[tbh]
\centering
\includegraphics[scale=0.4]{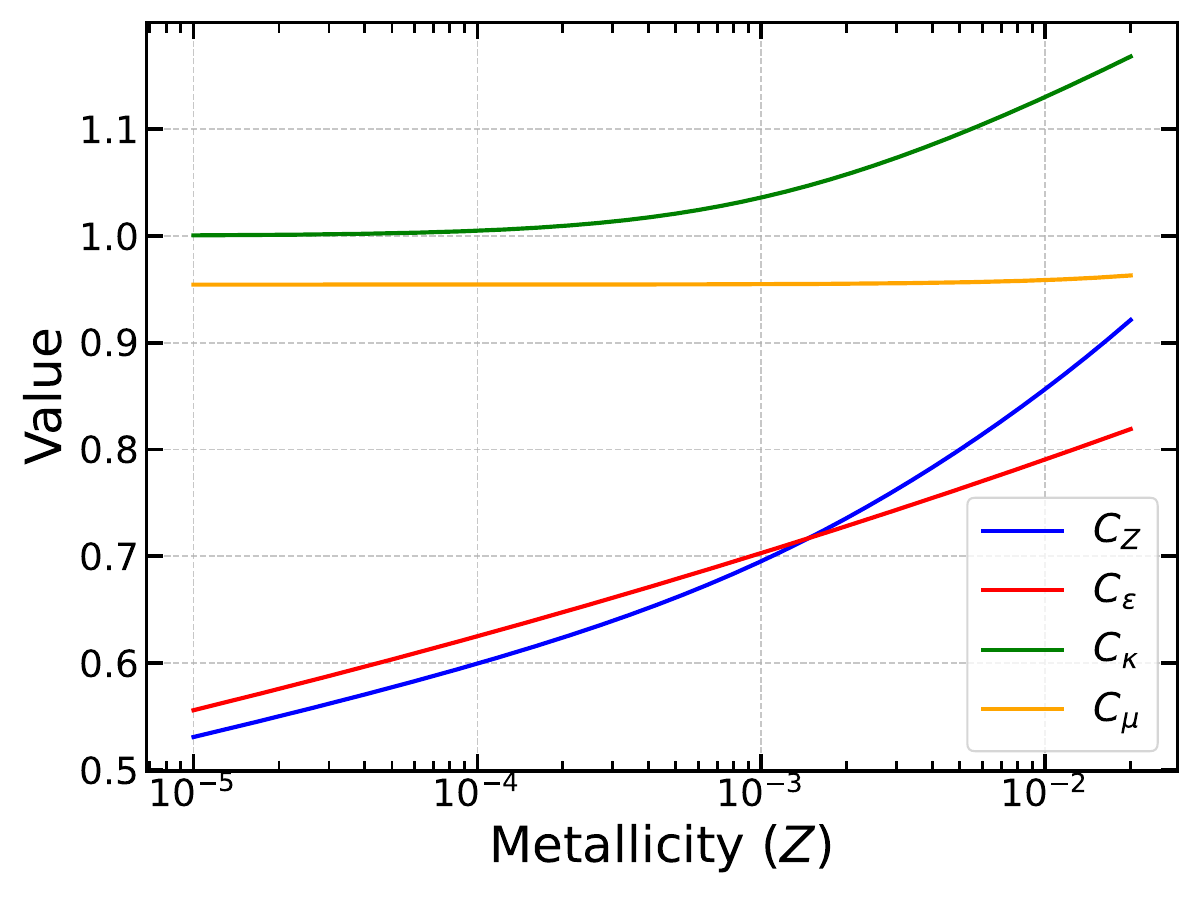}
\caption{The coefficient of radius $C_Z$ as a function of metallicity, along with the individual contributions from nuclear energy generation ($C_{\epsilon}$), opacity ($C_{\kappa}$), and mean molecular weight ($C_{\mu}$), as derived from homology relations.}
\label{fig:homology}
\end{figure}

Substituting the expressions for $\kappa$ and $\epsilon$ into Equation (7), we obtain:
\begin{equation}
    R^{5/2} \sim (\kappa_0 \epsilon_0)^{1/4} T^{\frac{\nu -7.5}{4}} M.
\end{equation}
To eliminate $T$, we use a scaling relation derived by combining mass conservation, hydrostatic equilibrium, and the ideal gas law:
\begin{equation}
    T\sim \frac{\mu M}{R}.
\end{equation}
Substituting this expression for $T$ into Equation (7) yields:
\begin{equation}
    R\sim (\kappa_0 \epsilon_0)^{\frac{2}{2\nu +5}} \mu^{\frac{2\nu-15}{2\nu+5}}M^{\frac{2\nu-7}{2\nu+5}}
\end{equation}
For $\nu = 17$ (CNO cycle), this relation becomes:
\begin{equation}
R\sim (\kappa_0 \epsilon_0)^{0.051} \mu^{0.49} M^{0.69}.
\end{equation}
The three quantities $\kappa_0$, $\epsilon_0$, and $\mu$ all depend on $Z$. 
We now express the $Z$-dependence of each term. For Kramers opacity, which includes both free-free and bound-free absorption, the coefficient can be approximated as
\begin{equation}
\kappa_0 \sim (1+10^3 Z). 
\end{equation}
When $Z \gg 10^{-3}$, bound-free absorption dominates and thus $\kappa_0 \sim Z$; when $Z \ll 10^{-3}$, free-free absorption dominates, and thus the opacity becomes largely independent of $Z$. Near the critical metallicity $Z\sim 0.001$ for RSG formation shown in the models of Paper I, both contributions are comparable. The coefficient $\epsilon_0$ for the burning of the CNO cycle is proportional to the mass fraction of carbon, nitrogen, and oxygen, and we approximate it as $\epsilon_0 \approx Z$. For a fully ionized ideal gas, the mean molecular weight is approximated by:
\begin{equation}
\mu \simeq \frac{4}{3+5X-Z}.
\end{equation}
Assuming a fixed He mass fraction $Y = 0.278$ in our models, the H mass fraction becomes $X = 0.722 - Z$, and thus:
\begin{equation}
    \mu^{-1}\simeq \frac{6.61-6Z}{4}.
\end{equation}
Substituting these $Z$ dependencies into Equation (11), we obtain:
\begin{equation}
R\sim [Z(1+10^3 Z)]^{0.051} (1.1-Z)^{-0.49} M^{0.69}.
\end{equation}
We define the coefficient dependent on $Z$ as $C_Z \equiv [Z(1+10^3 Z)]^{0.051} (1.1-Z)^{-0.49}$, with contributions from nuclear burning $C_{\epsilon}\equiv Z^{0.051}$, opacity $C_{\kappa}\equiv (1+10^3Z)^{0.051}$, and mean molecular weight $C_{\mu} = (1.1-Z)^{-0.49}$. The values of these coefficients as functions of $Z$ are plotted in Figure~\ref{fig:homology}.

Among the three components, the contribution from mean molecular weight is relatively mild, as $(1.1 - Z)^{-0.49}$ remains close to unity and varies only slightly across the metallicity range considered. For $Z \lesssim 10^{-3}$, the $C_{\kappa}$ also stays near unity, making $C_{\epsilon}$ the dominant contributor to $C_Z$. At higher metallicities ($Z \gtrsim 10^{-3}$), the contribution from opacity becomes more significant, while $C_{\epsilon}$ continues to play a role. Overall, the increase in stellar radius with metallicity during the main sequence is primarily due to enhanced opacity and more efficient energy generation through the CNO cycle.

At fixed metallicity, Equation (15) gives $R \propto M^{0.69}$. We compare this scaling with the threshold TAMS radius for RSG formation across different stellar masses, shown in Figure~\ref{fig:m_radius_TAMS}, which follows an approximate relation $R \propto M^{0.75}$. Although the latter exponent (0.75) is only a coarse fit, it is close to the value (0.69) expected at constant metallicity. This similarity indicates that the mass dependence of the threshold TAMS radius is closely aligned with the mass dependence of stellar radius at fixed metallicity. As a result, the near constancy of the critical metallicity at $Z \sim 0.001$ over a wide mass range shown in the models of Paper I can be naturally explained.

\section{Impact of TAMS Radius on Supergiant Expansion}\label{sec:core}

The remaining key question is how $R_{\rm TAMS}$ regulates the subsequent supergiant expansion, and establishes a threshold for RSG formation. To address this, we first use the $25\,M_{\odot}$ models in Grids (c) and (a) to show the key role of $R_{\rm TAMS}$, and then classify each model in the full Grid (a) to examine the evolution of stars with different masses.

\subsection{Role of TAMS Radius for $25\,M_{\odot}$ Stars}

\begin{figure}[tbh]
\centering
\includegraphics[scale=0.4]{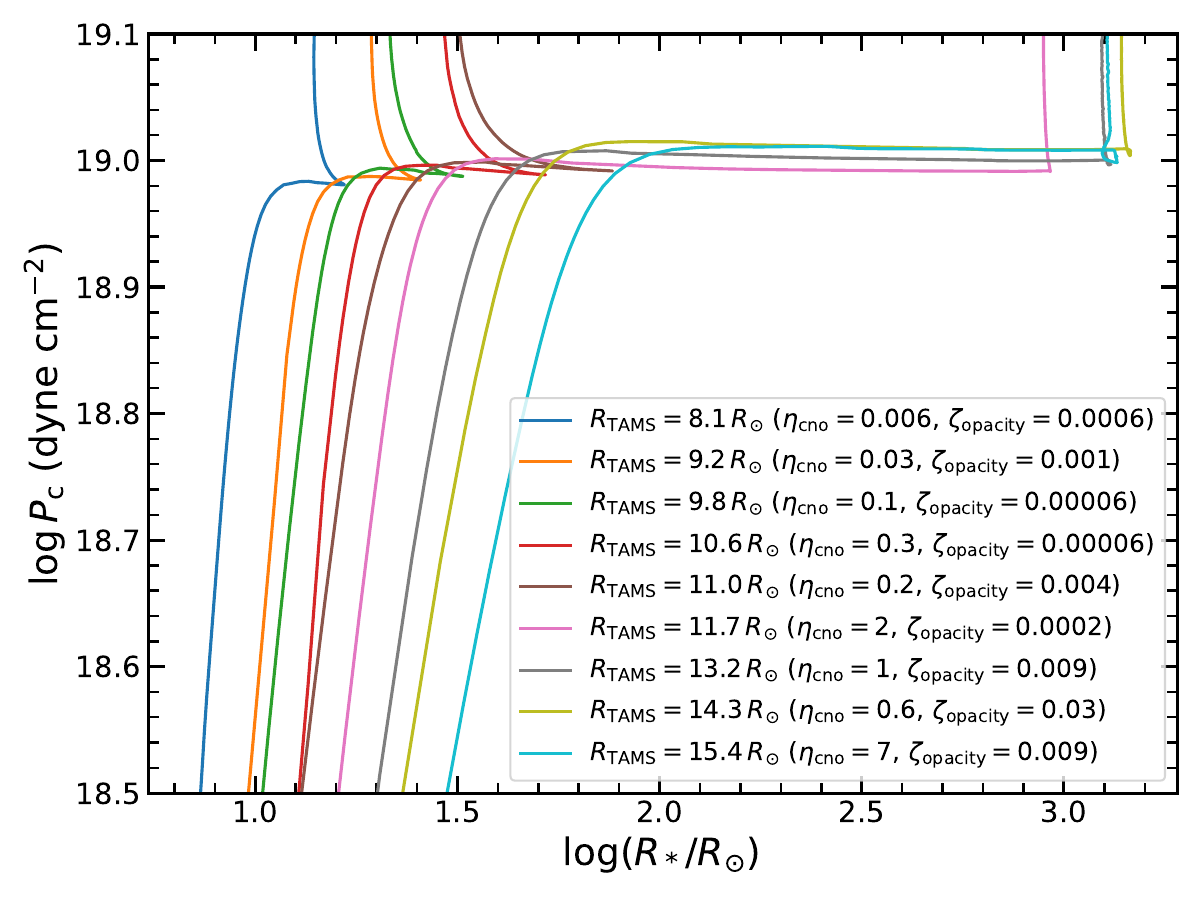}
\caption{Evolution of central pressure ($P_{\rm c}$) as a function of stellar radius ($R_*$) during the core He-burning stage for $25\,M_{\odot}$, $Z = 0.001$ models with varying $\eta_{\rm CNO}$ and $\zeta_{\kappa}$. The legend is ordered by the TAMS radius ($R_{\rm TAMS}$) of each model. Notably, the horizontal shifts in $R_*$ among the evolutionary tracks align with $R_{\rm TAMS}$, rather than with the individual input parameters $\eta_{\rm CNO}$ or $\zeta_{\kappa}$.}
\label{fig:central}
\end{figure}

Variations in $\eta_{\rm CNO}$ and $\zeta_{\kappa}$ in Grid (c) produce different values of $R_{\rm TAMS}$. To examine how $R_{\rm TAMS}$ influences later evolution, we select several example models from this grid for comparison. Figure~\ref{fig:central} shows their evolutionary tracks in the $(R_*, P_{\rm c})$ plane, where $R_*$ traces the envelope extent and $P_{\rm c}$ reflects the core conditions. All models follow a similar pattern in core evolution: $P_{\rm c}$ rises rapidly, enters a brief plateau, and then increases again. However, their tracks are systematically shifted along the $R_*$ axis, and this horizontal offset correlates strongly with $R_{\rm TAMS}$ rather than with the individual input parameters $\eta_{\rm CNO}$ or $\zeta_{\kappa}$. Thus, although all models undergo comparable core-evolution stages, those with larger $R_{\rm TAMS}$ reach each stage at correspondingly larger envelope radii.

\begin{figure}[tbh]
\centering
\includegraphics[scale=0.5]{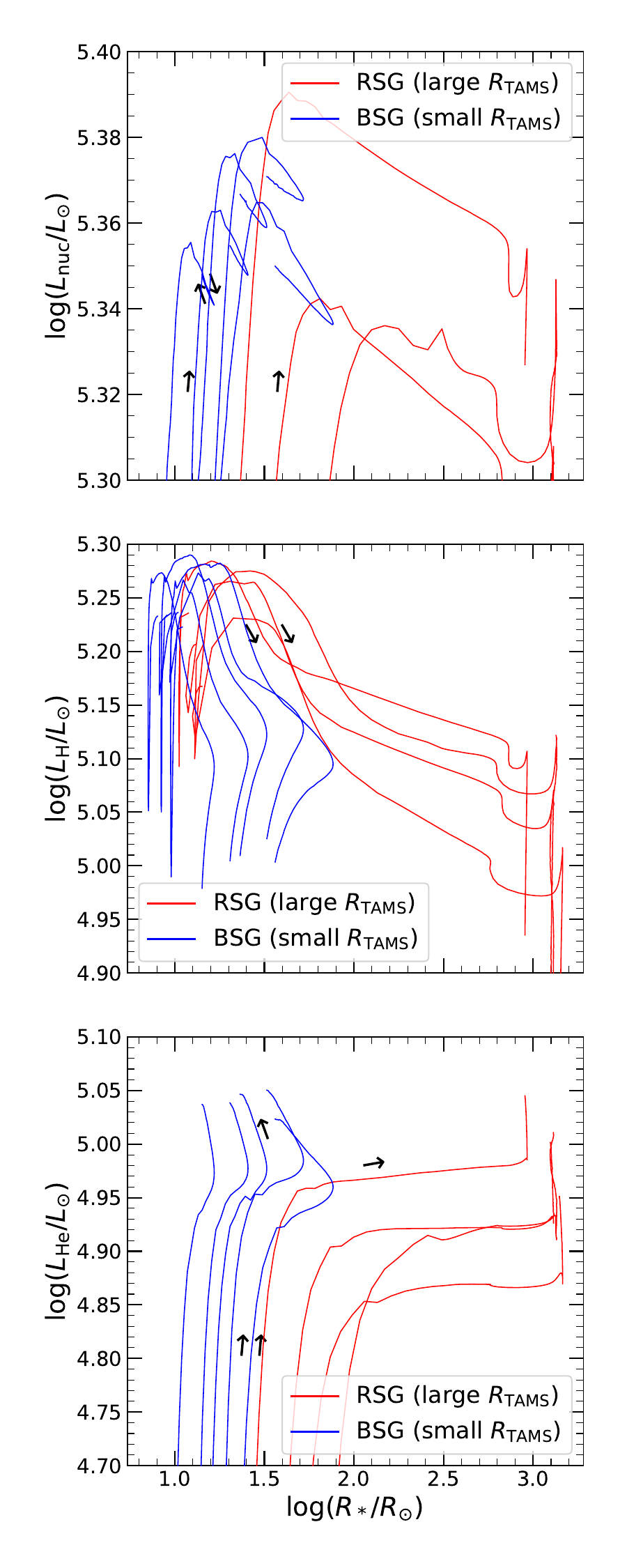}
\caption{Evolution of the nuclear luminosity $L_{\rm nuc}$ and its contributions from H burning ($L_{\rm H}$) and He burning ($L_{\rm He}$) as functions of stellar radius ($R_*$) for $25\,M_{\odot}$ models shown in Figure~\ref{fig:central}. Red curves denote models with larger $R_{\rm TAMS}$ that evolve into the RSG phase, while blue curves denote models with smaller $R_{\rm TAMS}$ that remain in the BSG phase. To reduce visual clutter, the tracks are truncated at the point where $L_{\rm He}$ reaches its maximum. Arrows indicate the direction of evolution.}
\label{fig:Lnuc}
\end{figure}

To explain how this systematic offset influences the fate of supergiants, we present the evolution of $L_{\rm nuc}$ for the same set of $25\,M_{\odot}$ models in Figure~\ref{fig:Lnuc}. For these models, the bifurcation into RSG or BSG outcomes follows the criterion of \citet{ou2024}: after core contraction ends, if $L_{\rm nuc}$ turns upward while the star is still in the BSG regime, envelope expansion halts and the model remains a BSG; conversely, if $L_{\rm nuc}$ continues to decline, the envelope expands into the RSG phase. Decomposing $L_{\rm nuc}$ into contributions from H burning ($L_{\rm H}$) and He burning ($L_{\rm He}$), we find that the key driver of the $L_{\rm nuc}$ upturn in BSG cases is the second rise of $L_{\rm He}$. This second rise occurs in all models. However, if it takes place only after the star has already expanded into the stable RSG regime, it has little influence on the envelope. In contrast, if the same increase occurs before the star enters the RSG phase, it can halt further expansion and cause the star to remain a BSG.

The crucial role of $R_{\rm TAMS}$ lies in setting the envelope radius at the evolutionary stage corresponding to the second rise of $L_{\rm He}$. A star with a larger $R_{\rm TAMS}$ maintains a systematically larger envelope radius at all subsequent core-evolution stages, including at the time of the second rise of $L_{\rm He}$. The timing of this second rise relative to the envelope expansion then determines whether the star ultimately ends its evolution as a BSG or an RSG.

\begin{figure}[tbh]
\centering
\includegraphics[scale=0.5]{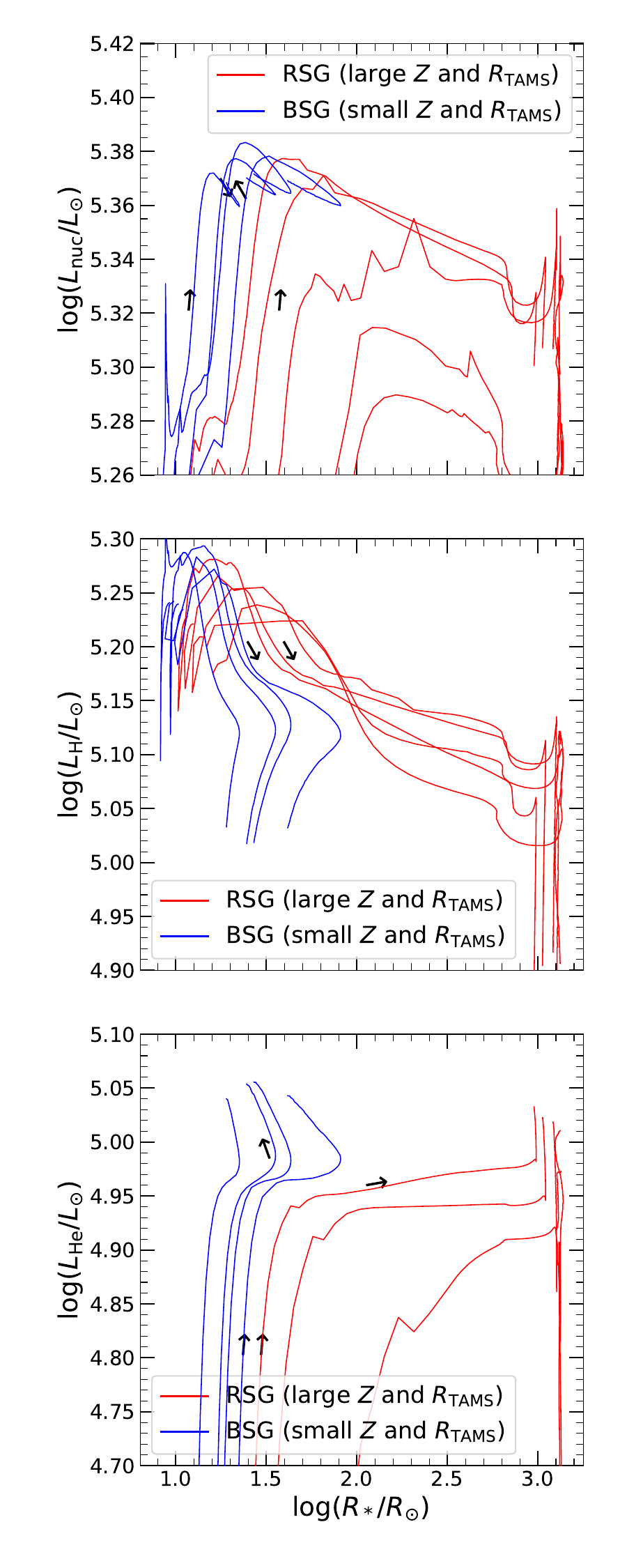}
\caption{Evolution of the nuclear luminosity $L_{\rm nuc}$ and its contributions from H burning ($L_{\rm H}$) and He burning ($L_{\rm He}$) as functions of stellar radius ($R_*$) for $25\,M_{\odot}$ models with different metallicities from Grid (a). Red curves denote higher-metallicity models (and thus larger $R_{\rm TAMS}$) that evolve into the RSG phase, while blue curves denote lower-metallicity models (with smaller $R_{\rm TAMS}$) that remain in the BSG phase. To reduce visual clutter, the tracks are truncated at the point where $L_{\rm He}$ reaches its maximum. Arrows indicate the direction of evolution.}
\label{fig:Lnuc_Z}
\end{figure}

We now examine the $25\,M_{\odot}$ models in Grid (a) across different metallicities by plotting the evolution of $L_{\rm nuc}$, $L_{\rm H}$, and $L_{\rm He}$ as functions of $R_*$, as shown in Figure~\ref{fig:Lnuc_Z}. Lower-metallicity stars have smaller $R_{\rm TAMS}$ and therefore encounter the second rise of $L_{\rm He}$ at an earlier evolutionary stage, which prevents them from expanding into the RSG regime. In contrast, higher-metallicity stars enter the RSG phase before the second rise of $L_{\rm He}$ occurs. This behavior is consistent with the mechanism identified in Grid (c) (Figure~\ref{fig:Lnuc}).


\subsection{Classification of Supergiant Evolution Pathways}

To develop a general explanation for the critical metallicity across different stellar masses, we examine the supergiant evolutionary pathways of all the models in Grid (a). We classify the models following \citet{ou2024}, in which we identified two routes for stars to enter the RG/RSG phase, both governed by the refined “mirror principle.” The first is direct envelope expansion during core contraction, referred to here as the \textbf{RSG-C} case. The second is continued expansion after core contraction ends, driven by the decline of the nuclear luminosity $L_{\rm nuc}$—the combined contribution of shell H burning and core He burning—which we call the \textbf{RSG-L} case. In contrast, if $L_{\rm nuc}$ rises again before turning into an RG/RSG, the envelope re-contracts and the star remains a BSG; we denote this as the \textbf{BSG-L} case.

In addition to the above cases, there exists an additional type for stars with initial masses $\gtrsim 30\,M_{\odot}$ and metallicities $Z \lesssim 0.001$ that the C core begins to form while the He core is still contracting. The envelope may or may not expand further after the onset of core C burning, and this complex evolution in the final stages lies beyond the scope of this study. We categorize these stars as \textbf{BSG-E} on the basis of their status before the formation of the C core.

We classify the core He-burning evolution of massive stars into four distinct types:
\begin{enumerate}
    \item[(1)] \textbf{RSG-C}: Core contraction directly leads to expansion into the RSG regime.
    \item[(2)] \textbf{RSG-L}: $L_{\rm nuc}$ undergoes a sustained decline after the core contraction, driving a continued expansion of the envelope in the RSG regime.
    \item[(3)] \textbf{BSG-L}: $L_{\rm nuc}$ increases shortly after core contraction, causing the envelope to contract, and the star remains a BSG.
    \item[(4)] \textbf{BSG-E}: Early C core formation occurs during ongoing He core contraction; the star remains in the BSG phase at this time.
\end{enumerate}
\textbf{The detailed classification criterion for each type is described in Appendix~\ref{sec:appA}.}

\subsection{TAMS Radius and the Critical Metallicity}

\begin{figure*}[tbh]
\centering
\includegraphics[scale=0.4]{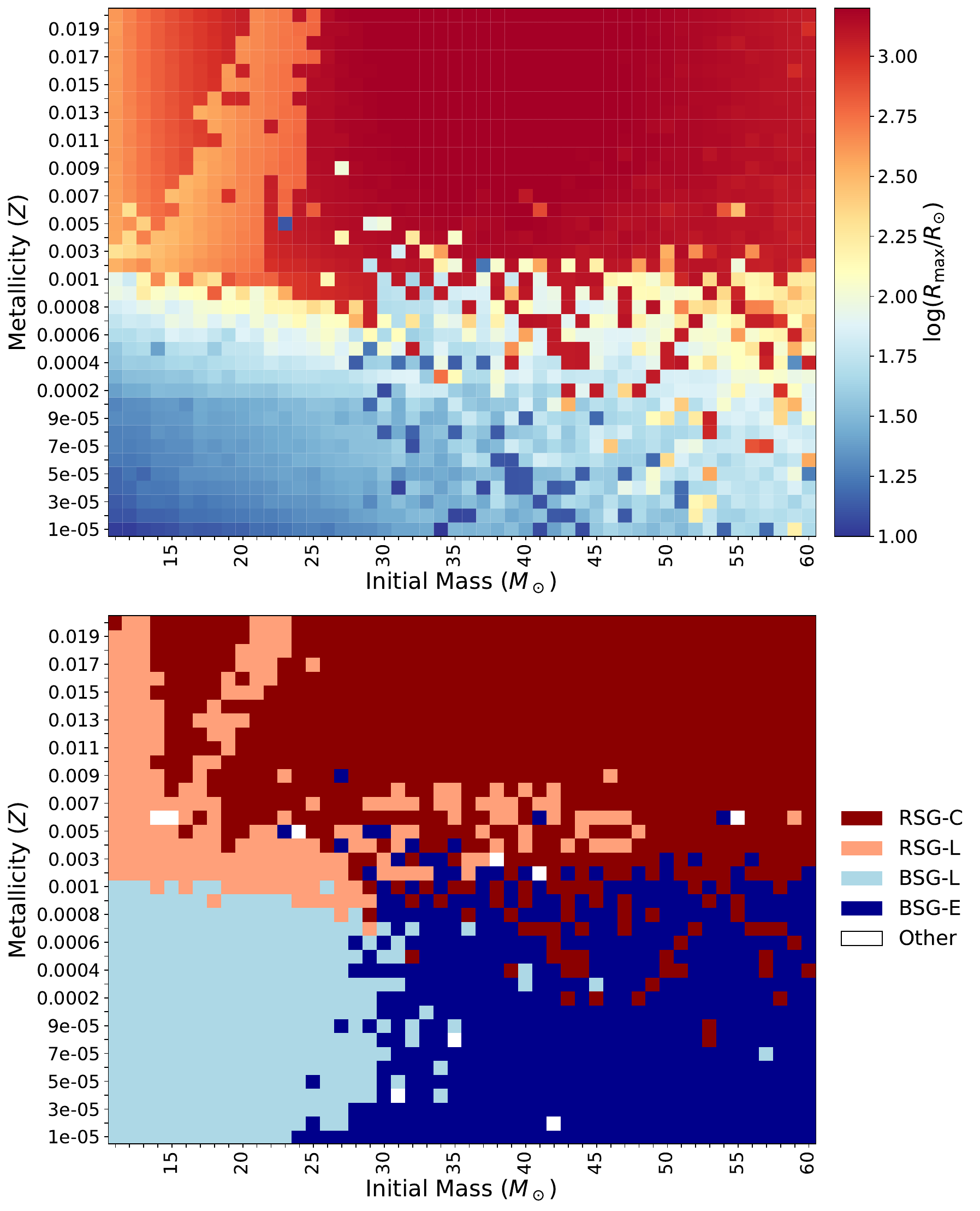}
\caption{\textit{Top panel:} Maximum stellar radius during the core He-burning stage ($R_{\rm max}$) for each model in Grid (a), spanning initial masses of $11$–$60\,M_{\odot}$ and metallicities of $Z = 10^{-5}$–$2 \times 10^{-2}$. 
\textit{Bottom panel:} Classification of each model's evolutionary pathway into these types: \textbf{RSG-C}, \textbf{RSG-L}, \textbf{BSG-L}, \textbf{BSG-E}, and Other. In general, stars with initial masses $\lesssim 30\,M_{\odot}$ and metallicities $Z \lesssim Z_{\rm c} \sim 0.001$ follow the \textbf{BSG-L} pathway, while more massive stars ($\gtrsim 30\,M_{\odot}$) with at similarly low metallicities are classified as \textbf{BSG-E}. Stars with $Z \gtrsim Z_{\rm c}$ are classified as either \textbf{RSG-C} or \textbf{RSG-L}.}
\label{fig:classification}
\end{figure*}

The classification results are shown in Figure~\ref{fig:classification} and compared with the maximum supergiant radii. The critical metallicity for cool-supergiant formation at $Z \sim 0.001$ shown in Grid (a) does not correspond to a single evolutionary boundary. Instead, it aligns with different pathway transitions in different mass ranges: it marks the boundary between the \textbf{RSG-L} and \textbf{BSG-L} cases for $M_i \lesssim 30\,M_{\odot}$, but the boundary between the \textbf{RSG-C} and \textbf{BSG-E} cases for $M_i \gtrsim 30\,M_{\odot}$.

The transition mechanism between the \textbf{RSG-L} and \textbf{BSG-L} cases for $M_i \lesssim 30\,M_{\odot}$ follows the explanation in Section 6.1, which used the $25\,M_{\odot}$ models as examples: $R_{\rm TAMS}$ is the decisive factor for RSG formation because it determines the envelope’s extent when $L_{\rm He}$ undergoes its second rise. Stars with higher metallicity—and therefore larger $R_{\rm TAMS}$—have already entered the RSG regime by the time of this second rise and thus remain RSGs. In contrast, low-metallicity stars are still in the BSG regime when the second rise in $L_{\rm He}$ occurs, preventing further expansion and causing them to remain BSGs. Their mass range $11$–$30\,M_{\odot}$ aligns with the classical red supergiant population \citep{meynet2000,beasor2025}, many of which are expected to end their lives as core-collapse supernovae \citep{smartt2009}.

The transition between \textbf{RSG-C} and \textbf{BSG-E} cases for stars with $M_i \gtrsim 30\,M_{\odot}$ can likewise be interpreted in terms of $R_{\rm TAMS}$. For models with smaller $R_{\rm TAMS}$, advanced core-evolution stages occur while the envelope is still relatively compact. The \textbf{BSG-E} cases are those in which the star proceeds into core C burning while the He core is still contracting and the envelope is still expanding in the BSG regime.

\subsection{Correspondence between core and envelope evolution}

\begin{figure*}[tbh]
\centering
\includegraphics[scale=0.3]{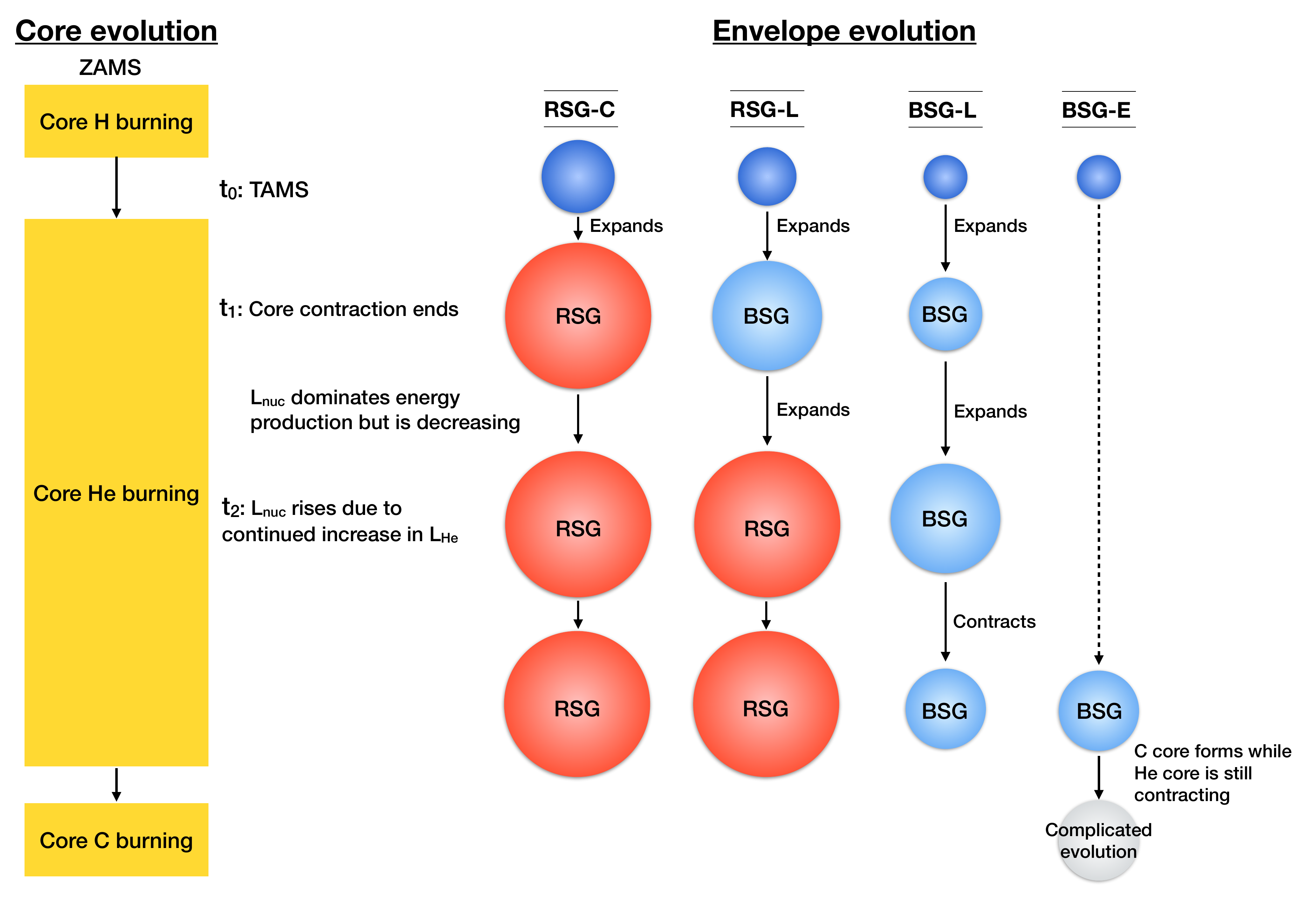}
\caption{Flow chart illustrating the evolutionary stages of the stellar core and envelope. The stellar radius at the TAMS is the key factor that determines the evolutionary pathway and the resulting supergiant type.}
\label{fig:flow}
\end{figure*}

The bifurcation between RSG and BSG outcomes arises from how the evolutionary stages of the core and the envelope correspond to one another. As summarized in the flow chart in Figure~\ref{fig:flow}, the post–main-sequence evolution of the core and the envelope is not synchronized; each follows its own sequence of structural transitions. After the TAMS, the core begins to contract. When core contraction ends, $L_{\rm nuc}$ first decreases and then increases again due to the second rise of $L_{\rm He}$. A more advanced core stage is reached once the C core forms. Meanwhile, the envelope initially expands into the BSG regime and may later continue expanding into the RSG regime.

The importance of $R_{\rm TAMS}$ is that it is the stellar radius at the onset of post–main-sequence evolution, serving as the initial condition for subsequent envelope expansion, and thus leading to different evolutionary pathways. A larger $R_{\rm TAMS}$ allows the star to reach large radii earlier during its expansion, so that it is already in the stable RSG state when the second rise of $L_{\rm He}$ occurs (the \textbf{RSG-C} and \textbf{RSG-L} cases). Conversely, a smaller $R_{\rm TAMS}$ keeps the star compact until a later core-evolution stage, causing the second rise of $L_{\rm He}$ to occur while the star is still in the BSG regime (\textbf{BSG-L} cases), or—at higher masses—to proceed quickly into core C burning while still compact (\textbf{BSG-E} cases).

\section{Discussions and Conclusions}\label{sec:conclusions}

This study investigates the physical origin of the critical metallicity for RSG formation through a series of numerical experiments. We show that metallicity determines the supergiant outcome primarily through its influence on the stellar radius. In particular, $R_{\rm TAMS}$—strongly shaped by metallicity—defines a threshold for RSG formation, with the threshold value depending on stellar mass.

Metallicity significantly influences stellar evolution through its effects on opacity, nuclear reaction rates, and the mean molecular weight. Using stellar models together with dimensional analysis, we show that higher-metallicity stars tend to reach larger $R_{\rm TAMS}$ due to enhanced opacity and increased CNO-cycle energy generation. The resulting $R_{\rm TAMS}$ then serves as the initial condition for envelope expansion during post–main-sequence evolution.

The mechanism linking $R_{\rm TAMS}$ to the supergiant outcome lies in the correspondence between the envelope radius and the core evolutionary stage. A larger $R_{\rm TAMS}$ means that post–main-sequence expansion begins from a larger initial radius, allowing the envelope to reach larger sizes at subsequent core-evolution phases. For stars with $M_{\rm i} \sim 11$–$30\,M_{\odot}$, $R_{\rm TAMS}$ determines whether the envelope has already expanded to RSG dimensions by the time of the second rise of $L_{\rm He}$, after which further expansion is suppressed. For more massive stars with $M_{\rm i} \sim 30$–$60\,M_{\odot}$, a smaller $R_{\rm TAMS}$ can lead the star to enter core carbon burning while the envelope remains compact in the BSG phase. The evolution in this higher-mass regime is more complex and requires further investigation.

In addition to metallicity, other stellar properties—such as rotation and mass loss—have been suggested to influence whether a star evolves into a red or blue supergiant \citep[e.g.,][]{He2025}. Our results imply that any parameter capable of modifying $R_{\rm TAMS}$ can affect the supergiant outcome. Consequently, one may expect not only a critical metallicity but also threshold values in other physical parameters for RSG formation, which will be explored in future studies.

It is worth noting that the supergiant outcome is sensitive to mixing prescriptions in stellar models \citep[e.g.,][]{langer1995,sibony2023}. In particular, recent studies by \citet{Schootemeijer2019} showed that enhanced overshooting and reduced semi-convection tend to favor evolution toward RSGs rather than BSGs. To assess the impact of these mixing parameters, we present additional grids of models in Appendix~\ref{sec:appB}, in which the overshooting parameter $f_{\rm ov}$ and the semi-convection efficiency $\alpha_{\rm sc}$ are modified relative to Grid (a). We find that while $f_{\rm ov}$ and $\alpha_{\rm sc}$ influence the detailed supergiant population, the critical metallicity at $Z \sim \text{a few} \times 10^{-3}$ is largely unchanged. Furthermore, for certain values of the convective overshooting parameter, stellar evolutionary tracks may undergo a ``blue loop" after entering the RSG phase \citep[e.g.,][]{wagle2019,wagle2020}. This behavior is beyond the scope of the present study and warrants further investigation to obtain a more complete picture of BSG/RSG evolution.

From stellar models, we have identified the physical mechanism that causes low-metallicity massive stars to remain blue rather than evolving into the RSG phase. These distinct evolutionary pathways at low metallicity have important implications for the nature of stellar populations and feedback in the early universe. For example, since low-metallicity stars are less likely to evolve into RSGs, a major progenitor class for core-collapse supernovae, the resulting supernova types in metal-poor galaxies may differ significantly from those observed in the local universe. Future observations targeting galaxies with metallicities as low as $\sim 0.1\,Z_{\odot}$ will be crucial in testing these predictions.

\begin{acknowledgments}
This research is supported by the National Science and Technology Council, Taiwan, under grant No. MOST 110-2112-M-001-068-MY3, NSTC 113-2112-M-001-028-, 114-2112-M-001-012, 114-2811-M-001-094, and Academia Sinica, Taiwan, under a career development award under grant No. AS-CDA-111-M04. This research was supported in part by grant NSF PHY-2309135 to the Kavli Institute for Theoretical Physics (KITP) and grant NSF PHY-2210452 to the Aspen Center for Physics. KC acknowledges the support of the Alexander von Humboldt Foundation.
Our computing resources were supported by the National Energy Research Scientific Computing Center (NERSC), a U.S. Department of Energy Office of Science User Facility operated under Contract No. DE-AC02-05CH11231 and the TIARA Cluster at the Academia Sinica Institute of Astronomy and Astrophysics (ASIAA).

\end{acknowledgments}

\software{MESA \citep{paxton2011,paxton2013,paxton2015,paxton2018,paxton2019,jermyn2023}
          }

\clearpage
\appendix
\section{Classification of evolutionary pathways}\label{sec:appA}
To perform the classification, we examine the evolutionary tracks of each model using the approach described in Section 4 of \citet{ou2024}. We track the evolution of the stellar radius ($R_*$), central pressure ($P_{\rm c}$), and the nuclear energy generation rate ($L_{\rm nuc}$) from the formation of the He core to the formation of the C core. Within this time interval, we identify the maximum stellar radius ($R_{\rm max}$) and the maximum central pressure ($P_{\rm max}$), along with the corresponding stellar radius at the time of $P_{\rm max}$, denoted $R_{P\rm max}$.

If $R_{P\rm max}$ and $R_{\rm max}$ differ by less than 1\%, this indicates that the maximum radius is reached while the central pressure is still increasing, which implies that the core is still contracting at that time. For stars that satisfy this condition and have an effective temperature $\log (T_{\rm eff}/{\rm K}) < 3.8$ at $R_{\rm max}$, we classify them as \textbf{RSG-C}.

If $R_{P\rm max}$ occurs significantly earlier than $R_{\rm max}$, it indicates that the envelope continues to expand after the core contraction ends. In such cases, we identify the minimum $L_{\rm nuc}$ following $R_{P\rm max}$ and define the corresponding stellar radius at this time as $R_{L\rm min}$. If $R_{L\rm min}$ and $R_{\rm max}$ differ by less than 5\%, we classify the model as \textbf{RSG-L} if $\log (T_{\rm eff}/{\rm K}) < 3.8$ at $R_{\rm max}$, and as \textbf{BSG-L} if $\log (T_{\rm eff}/{\rm K}) \geq 3.8$ at $R_{\rm max}$.

For low-metallicity models, we identify 60 cases where the star remains a BSG throughout core He burning and exhibits $R_{P\rm max} \sim R_{\rm max}$. These cases cannot be reliably classified using automated criteria. We therefore assign classifications manually based on the evolution of $R_*$ and $L_{\rm nuc}$. Models that exhibit envelope contraction when $L_{\rm nuc}$ reaches the minimum and subsequently rises are classified as \textbf{BSG-L}, while those that continue expanding at the time of C core formation are classified as \textbf{BSG-E}.

\section{Effects of Convective Overshoot and Semi-convection}\label{sec:appB}

\begin{figure}[tbh]
\centering
\includegraphics[scale=0.4]{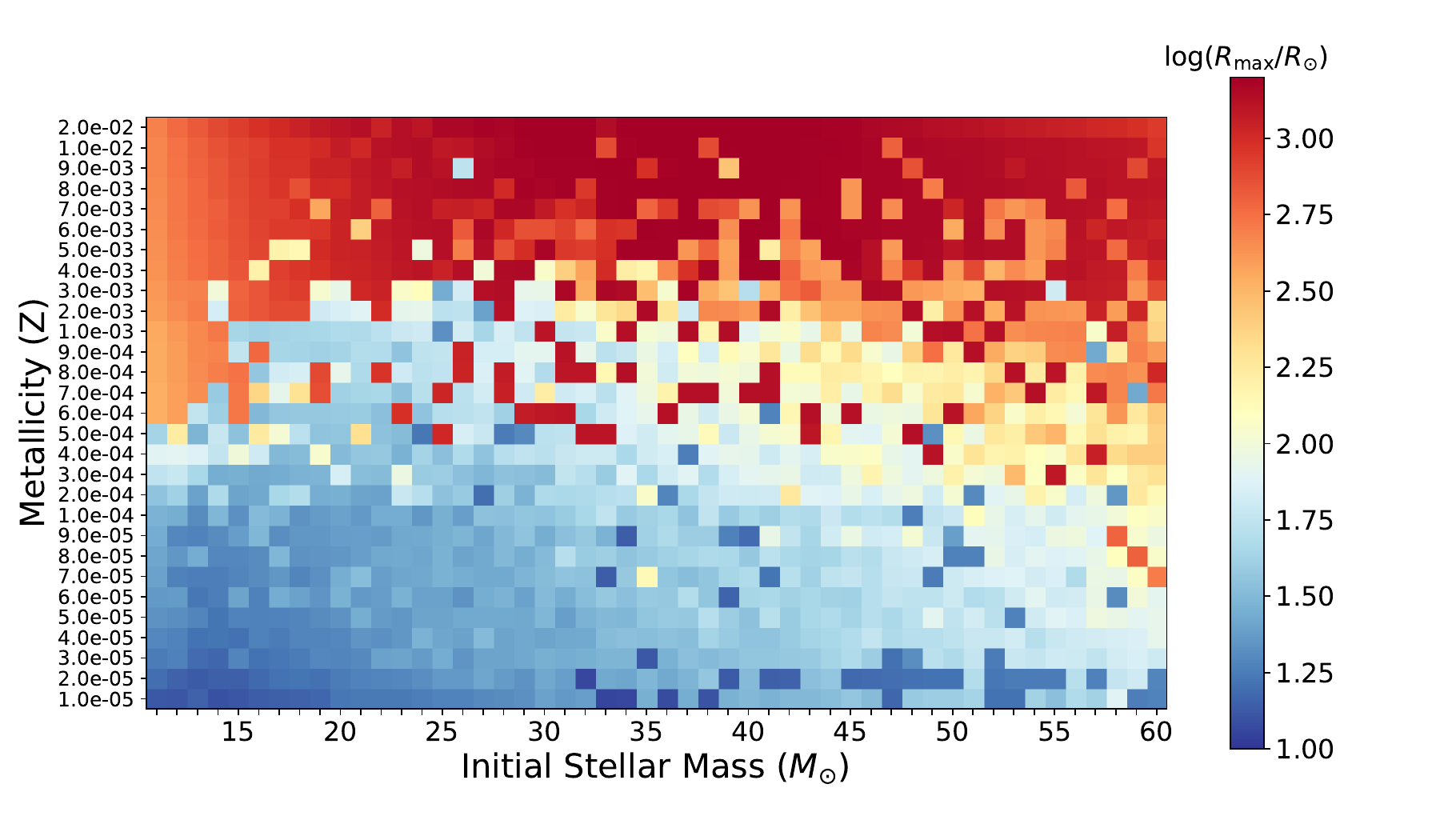}
\caption{Maximum stellar radius during the core He-burning phase ($R_{\rm max}$) for models with enhanced overshooting ($f_{\rm ov} = 0.01$). All other model parameters are identical to those of Grid (a).}
\label{fig:overshoot}
\end{figure}

In the main text, we use Grid (a) to demonstrate the critical metallicity for RSG formation, adopting an exponential overshooting prescription \citep{Herwig2000} with $f_{\rm ov} = 0.001$ in non-burning and H-burning regions and a semi-convection parameter of $\alpha_{\rm sc} = 0.01$. To test the sensitivity of the critical metallicity to overshooting, we increase the overshooting parameter by a factor of 10 to $f_{\rm ov} = 0.01$, and the resulting $R_{\rm max}$ values are shown in Figure~\ref{fig:overshoot}. Although increased scatter appears near the RSG–BSG boundary, the critical metallicity remains close to $Z \sim 0.001$.

\begin{figure}[tbh]
\centering
\includegraphics[scale=0.4]{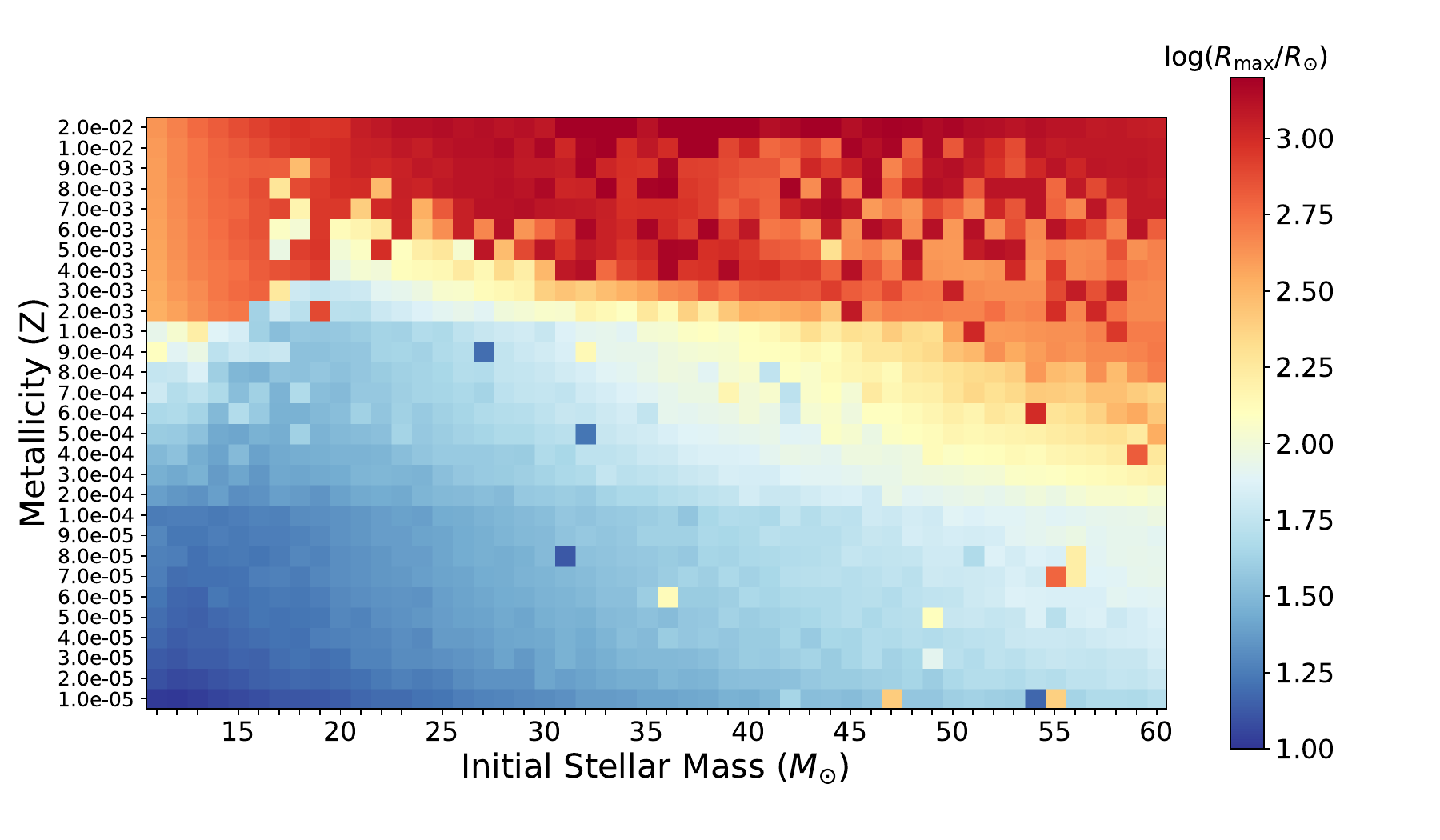}
\caption{Maximum stellar radius during the core He-burning phase ($R_{\rm max}$) for models with enhanced semi-convection efficiency ($\alpha_{\rm sc} = 1.0$). All other model parameters are identical to those of Grid (a).}
\label{fig:semiconv}
\end{figure}

We also examine the effect of semi-convection by increasing $\alpha_{\rm sc}$ by a factor of 100. As shown in Figure~\ref{fig:semiconv}, for $\alpha_{\rm sc} = 1.0$ the boundary between BSGs and RSGs shifts slightly upward at $M_{\rm i} \sim 20$–$30\,M_{\odot}$ and becomes smoother for $M_{\rm i} \gtrsim 50\,M_{\odot}$. Nevertheless, the overall trend of a critical metallicity at $Z \sim \text{a few} \times 10^{-3}$ persists.

\begin{figure}[tbh]
\centering
\includegraphics[scale=0.4]{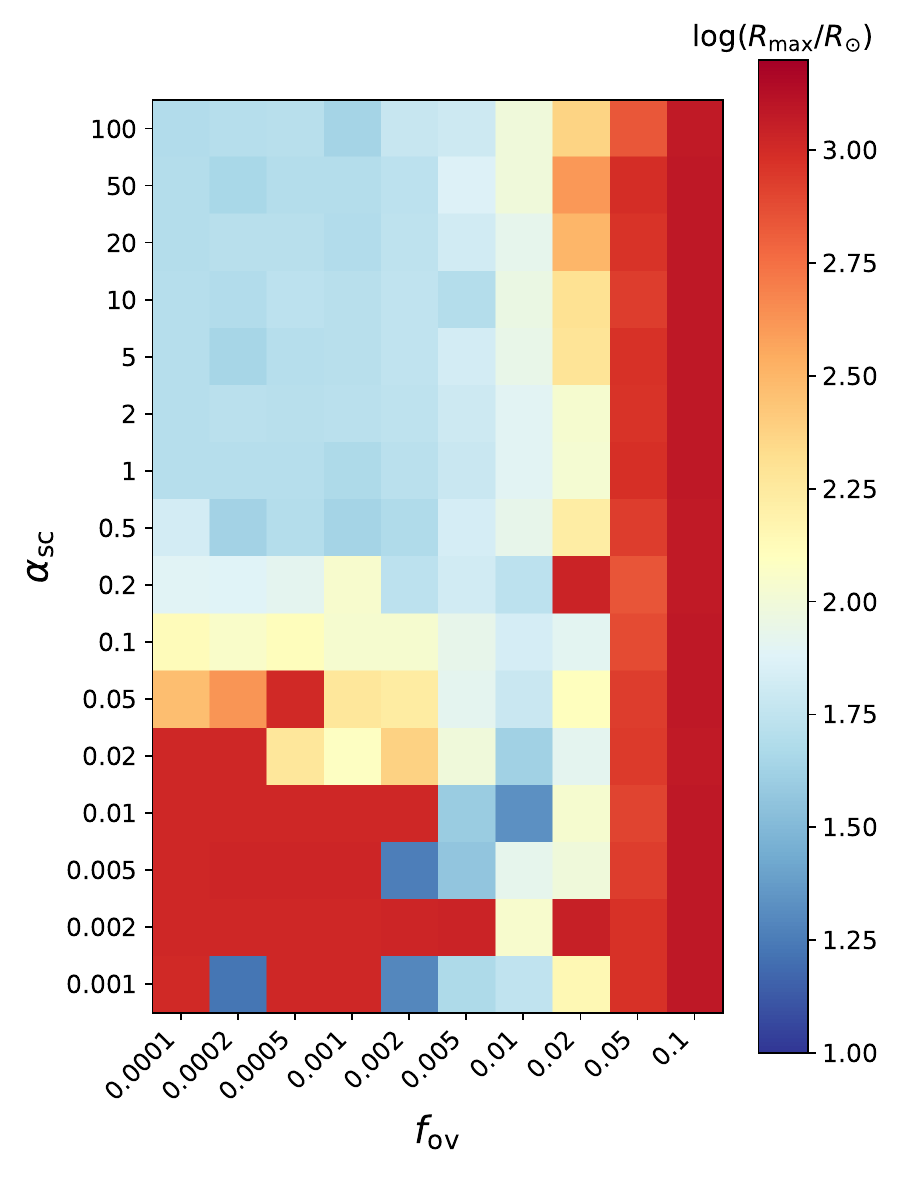}
\caption{Maximum stellar radius during the core He-burning stage ($R_{\rm max}$) for models with an initial mass of $25\,M_{\odot}$ and metallicity $Z=0.001$, computed for varying overshooting parameter $f_{\rm ov}$ and semi-convection parameter $\alpha_{\rm sc}$.}
\label{fig:ov-sc}
\end{figure}

Figure~\ref{fig:ov-sc} shows the resulting $R_{\rm max}$ for models with an initial mass of $25\,M_{\odot}$ and metallicity $Z=0.001$, varying the mixing parameters $f_{\rm ov}$ and $\alpha_{\rm sc}$. In general, these results are consistent with \citet{Schootemeijer2019}, who revealed that increased overshooting and reduced semi-convection promote evolution into the RSG phase.

\end{CJK*}

\end{document}